\documentclass{aa}
\usepackage{graphicx}
\def\ltsim{\mathrel{<\kern-1.0em\lower0.9ex\hbox{$\sim$}}}
\def\gtsim{\mathrel{>\kern-1.0em\lower0.9ex\hbox{$\sim$}}}
\def\udier{\"u}

\begin{document}
\title{Multi-frequency VLBA observations of compact sources from the 
               Peacock \& Wall catalogue}


\author{A. Rossetti \inst{1} \and
        F. Mantovani \inst{1} \and
         D. Dallacasa \inst{1,2} \and
        C. Fanti \inst{1,3} \and
        R. Fanti \inst{1,3} }
\offprints{A. Rossetti,\\
 \email{rossetti@ira.cnr.it}}

\institute{
Istituto di Radioastronomia del CNR, via Gobetti 101, I-40129, Bologna, Italy \and Dipartimento di Astronomia, Universit\`a degli Studi, via Ranzani 1, I--40127 Bologna, Italy \and Dipartimento di Fisica, Universit\`a degli Studi, via Irnerio 46, I--40126 Bologna, Italy}

\date{Received \today  / accepted}

\titlerunning{VLBA observations of PW compact sources}
        \authorrunning{A. Rossetti et al.}
   	
\abstract{VLBA observations are presented for 6 compact radio sources 
selected from the Peacock \& Wall catalogue. From the new morphological 
and spectral information 2 objects that in the Peacock and Wall catalogue 
are flat spectrum ($\alpha \le 0.5$) sources, appear to be double
sided objects with linear sizes of the order of one~kpc. Three are core-jet
sources and the last one is still an ``enigmatic'' object.
These data complete the sample of small double compact sources 
in the Peacock \& Wall catalogue and the complete list
is given.
}

\maketitle

\section{Introduction}

The Peacock \& Wall catalogue (Peacock \& Wall \cite{PW81}, here after referred
to as PW) lists radio sources in the Northern sky ($\delta > 10\degr$,
$|b|>10\degr$) stronger than 1.5~Jy at
2.7~GHz. This is  equivalent, for  a spectral index $\alpha = 0.7$ ({$S(\nu)
\propto \nu^{-\alpha}$}), to the flux density limit of the 3CRR catalogue at 
0.178~GHz (Laing et al. \cite{Laing83}). Due to the high frequency selection, the PW
catalogue contains a large fraction of sources with spectral index $\alpha_{0.178}^{2.7} \le$
0.5, many of which are missed in the 3CRR catalogue. 
A large number of them are genuine flat spectrum sources showing flux density 
variability and \emph{core-jet} radio structure on scales from tens of pc to kpc
\footnote{H$_{0}$=100~h~km~s$^{-1}$~Mpc$^{-1}$, q$_{0}$=0.5.} and beyond.
In a number of cases, instead, the spectrum is flat
at $\approx$ 2~GHz, inverted at lower frequencies and steepens at higher 
frequencies, while the radio structure is on sub-kpc scales. These 
sources, called GHz Peaked-Spectrum (GPS) sources, are 
under-represented in low frequency selected samples and their difference 
from the ``truly flat spectrum compact sources'' was emphasized by the 
extensors of the catalogue themselves (Peacock \& Wall \cite{PW82}).

GPSs form a continuous class of sources with the Compact Steep-Spectrum (CSS) sources, sizes up to 20~kpc 
and  transparent synchrotron spectrum flattening below a few hundred MHz).
Indeed, it has been shown (Fanti et al. \cite{Fanti90} and O'Dea \& Baum \cite {O'Dea97}) that, when
plotted on a ``rest-frame turnover frequency -- linear size '' ($\nu_{\rm turn}-LS$) diagram,
GPSs and CSSs form a continuous sequence and show that the turnover frequency 
is a fair estimator of their total linear size.

 A complete sample of GPSs with structural and polarization information has been presented
by Stanghellini et al. (\cite{Stanghellini98}). For a review on CSS/GPS properties see O'Dea (\cite{O'Dea98}).

A large fraction of CSSs and GPSs have a two-sided radio structure with
respect to the radio core. They are named CSOs (Compact Symmetric Objects) and MSOs 
(Medium Symmetric Objects) according to their
sub-kpc or over-kpc size. These sources are 
an important tool to investigate radio source evolution, since 
they are now interpreted as the young phase of the powerful large size
(hundreds of kpc) radio sources (see, e.g., Phillips \& Mutel \cite{Phillips82}; 
 Carvalho \cite{Carvalho85}; Fanti et al. \cite{Fanti95}; 
Readhead et al. \cite{Readhead96b}, Snellen et al. \cite{Snellen00}, \cite{Snellen03}).
The expansion velocities recently measured for a dozen CSOs 
(Owsianik \& Conway \cite{Owsianik98}; Polatidis \& Conway \cite{Polatidis03}) and the determination 
of the radiative ages of a large number of MSOs (Murgia et al. \cite{Murgia99}, 
Murgia \cite{Murgia03}) have provided strong support for this scenario.

A sample of CSOs and MSOs suitable for statistical studies
selected from the PW catalogue with the
criterion $\alpha^{5.0}_{2.7} \gtsim 0.5$ has been studied 
by Spencer et al. (\cite{Spencer89}), Fanti et al. (\cite{Fanti90}) and 
Dallacasa et al. (\cite{Dallacasa95}, \cite{Dallacasa04}).
The spectral selection they applied, while selecting CSOs and MSOs which peak below 
about 2~GHz (hence expected
to be typically larger than few hundred pc from the $\nu_{\rm turn}$--$LS$ plot),
would miss smaller size sources
with turnover frequency between 2 and 5~GHz. These sources would remain
hidden among the truly flat spectrum core-jet sources and can be recognized 
only by means of high resolution observations. 
Several PW sources with $\alpha \le 0.5$ were already imaged with the 
appropriate resolution and showed up as CSOs (for instance 
\object{0710+439} and \object{2352+495}, Wilkinson et
al. \cite{Wilkinson94}; \object{OQ208} Stanghellini et al.
 \cite{Stanghellini97}), but, at the time we started 
this investigation, for six of them the available radio structure 
information did not allow, in our opinion, a secure radio morphological 
classification.

In this paper, we present the results of the VLBA observations of 
the 6 mentioned radio sources.

The source list and observations are presented in Section 2, while in 
Section 3 we describe  the data reduction procedures. In Section 4 we 
discuss the morphological and spectral properties of each source. 
In Section 5 we report the complete list of CSOs/MSOs from the PW
catalogue, with the relevant data.
Conclusions are given in Section 6.

\section{The sample}

In order to find new CSSs/GPSs within the PW flat spectrum population, we 
searched in the literature 
for both sub-arcsecond structure information (Fey et al. \cite{Fey96}, 
\cite{Fey97}; Xu et al. \cite{Xu95}; Kellermann et al. \cite{Kellermann98}) 
and spectral shape (K\udier hr et al. \cite{Kuhr81}; Herbig \& Readhead 
\cite{HR92}). 
For a large fraction of these flat spectrum sources 
they were available and allowed a firm classification as core-jet sources.

\begin{table}[htbp]
\centering
 \caption{General source parameters:
Column 1: Source name;
Column 2: Optical identification (Id), G = galaxy, Q = quasar;
Column 3: Redshift (z) from Johnston et al. \cite{Johnston95};
Column 4: V magnitude ($m_\mathrm{v}$) from Johnston et al. \cite{Johnston95};
Column 5: Flux density at 5~GHz from Johnston et al. \cite{Johnston95};
Column 6: Flux density at 8.4~GHz from Patnaik et al. \cite{Patnaik92};
Column 7: Linear Size (LS) from previous VLBI observations.}
 \begin{tabular}{lllllll}
 \hline
 \hline
 Source            & Id &       z        & $m_\mathrm{v}$ &    $S_{5}$      &     $S_{8.4}$    &  LS     \\
                   &    &                &                &      Jy         &        Jy        &  pc     \\
   (1)             & (2)&      (3)       &    (4)         &      (5)        &        (6)       &  (7)    \\
\hline
 \object{0133+476} &  Q &  0.859         &    19.0        &     3.3         &    1.70          &$>15$    \\
 \object{0153+744} &  Q &  2.338         &    16.0        &     1.5         &    0.91          &$\sim12$ \\
 \object{0202+149} &   ?&  0.405$^{(1)}$ &    21.9        &     2.4         &    2.27$^{(2)}$  &$\sim6$  \\
 \object{0859+470} &  Q &  1.462         &    18.7        &     1.8         &    0.96          & $<3$    \\
 \object{0945+664} &  G &  --            &    21.6        &     1.4$^{(3)}$ &    0.78          &   --    \\
 \object{0954+556} &  Q &  0.909         &    17.7        &     2.3         &    1.55          &   --    \\
 \hline
 \end{tabular}
\vspace {0.10cm}
\begin{list}{}{}{}
{\scriptsize
\item[${(1)}$]Perlman et al. \cite{Perlman98}, AJ, 115, 1253
\vspace {0.10cm}
\item[${(2)}$]Wright \& Otrupcek, PKS Catalogue \cite{WO90}
\item[${(3)}$]Becker et al. \cite{Becker91}, ApJS, 75, 1
}
\end{list}
\label{parameters}
\end{table} 

However, for 6 objects we could not find any information or the existing
images did not  allow, in our opinion, a secure classification.
In some of these sources flux density measurements showed a moderate steepening 
of the radio spectrum above 5~GHz, while the others showed a flat spectrum
over the whole explored range of frequencies.
 
The sources are listed in Table ~\ref{parameters}. Four of them are
associated with quasars and one is a radio galaxy. The last one (\object{0202+149}) 
is still an ``enigmatic'' object (see Section 4).

\section{Observations and data reduction}

The VLBA observations were carried out in 4 different runs, from 
May 1999 to November 1999 (see Table ~\ref{observation}). 
\tabcolsep 0.06cm
\begin{table}[htbp]
\centering
 \caption{Observation data:
Column 1: Source name;
Column 2: Frequency;
Column 3: Date of observation;
Column 4: Missing antennas (empty = 10 VLBA antennas);
Column 5: Shortest projected baseline.
Column 6: Beam angular size of major, minor axis and position 
angle of the major axis;
Column 7: rms noise level measured on the images.
}
 \scriptsize
 \begin{tabular}{lrccccccc}
 \hline
 \hline
Source            & $\nu$ & Obs date & missing &u$_\mathrm {min}$&       &  beam &       & noise     \\
                  & GHz   &          & antennas&    M$\lambda$   &  mas  &  mas  &$\degr$&  mJy/beam \\
      (1)         & (2)   & (3)      &   (4)   &           (5)   & ---   &   (6) &  ---  &    (7)    \\
\hline
\object{0133+476} & 1.7   &02/05/1999& NL      &        1.1      &  6.7  &   4.0 &  -3   &  0.12     \\
                  & 4.8   &02/05/1999& NL      &        2.5      &  2.6  &   2.0 &  -8   &  0.07     \\
                  & 8.4   &20/06/1999& MK      &        4.0      &  1.5  &   1.1 &  24   &  0.20     \\
\object{0153+744} & 4.8   &02/05/1999& NL      &        2.5      &  1.3  &   1.2 &  78   &  0.13     \\
                  & 15    &20/06/1999& MK      &        8.3      &  0.7  &   0.5 & -64   &  0.08     \\
\object{0202+149} & 8.4   &20/06/1999& MK      &        5.1      &  2.4  &   1.4 & -15   &  0.12     \\
                  & 15    &20/06/1999& MK      &        9.2      &  1.1  &   0.6 & -22   &  0.35     \\
\object{0859+470} & 15    &13/11/1999&         &        9.5      &  0.9  &   0.6 & -16   &  0.09     \\
\object{0945+664} & 1.7   &03/11/1999&         &        0.8      &  7.2  &   6.3 &  -7   &  0.13     \\
                  & 4.8   &03/11/1999&         &        2.4      &  18   &   13  &  80   &  0.38     \\
                  & 15    &13/11/1999&         &        7.5      &  0.8  &   0.6 &  -8   &  0.06     \\
\object{0954+556} & 1.7   &03/11/1999&         &        1.1      &  7.3  &   4.3 &  12   &  0.13     \\
                  & 4.8   &03/11/1999&         &        2.5      &  2.1  &   1.8 & -18   &  0.08     \\
                  & 15    &13/11/1999&         &        7.1      &  0.9  &   0.6 &  -4   &  0.06     \\
 \hline
 \end{tabular}
 \label{observation}
\end{table}

\tabcolsep 0.08cm
\begin{table*}[htbp]
\centering
 \caption{Observed source parameters derived from images in Fig. ~\ref{images}:
Column 1: source name;
Column 2: observing frequency (GHz);
Column 3: source total flux density (Jy) measured by using TVSTAT;
Columns 4,5 : Largest Angular Size (LAS, mas) and Largest Linear Size (LLS, pc) 
(for the galaxy \object{0945+664} we assumed z$=1$); 
Column 6: sub-component label; components with an ``$^{*}$'' are extended; 
Columns 7,8: distance r (mas) and P.A. (\degr) from the reference component;
Column 9: deconvolved angular size of major and 
minor axis (mas) and the major axis position angle (degrees) from gaussian fit; 
for extended components linear sizes are derived from the images;
Column 10, 11: peak flux density (mJy/beam) and integrated flux density (mJy);
Column 12, 13: low and high frequency spectral indices.}
 \begin{tabular}{ccrrccccrrcrrrr}
\hline
 \hline
    Source       &$\nu$&$S_\mathrm{tot}$& LAS & LLS &  Comp   &      r     & P.A.&  Maj  & Min &  P.A. &$S_\mathrm{p}$ & $S_\mathrm{t}$&$\alpha_{lo}$&$\alpha_{hi}$\\
                 & GHz &        Jy      & mas &~pc~h$^{-1}$&  &     mas    &$\degr$&mas  & mas &$\degr$&       mJy/b   &       mJy     &             &             \\
        (1)      & (2) &        (3)     &(4)  &(5)  &   (6)   &     (7)    & (8) &  ---  & (9) &  ---  &     (10)      &     (11)      &     (12)    &     (13)    \\
 \hline 
\object{0133+476}& 1.7 &     1.39       &  33 & 137 &   A     &            &     &  3.37 & 1.43&  157  &     1080.4    &     1276.9    & -0.37       &   -0.55     \\
                 &     &                &     &     &Ext$^{*}$&            &     &   50  & 21  &       &       --      &      116.3    &             &             \\
                 & 4.8 &     1.88       &  20 &  83 &   A     &            &     &  0.95 & 0.29&  150  &     1678.3    &     1800.4    &             &             \\
                 &     &                &     &     &Ext$^{*}$&            &     &   27  & 12  &       &       --      &       79.6    &             &             \\
                 & 8.4 &     2.58       &   4 &  17 &   A1    &     0      &  0  &  0.47 & 0.12&  149  &     2286.4    &     2474.6    &             &             \\
                 &     &                &     &     &   A2    &2.5$\pm0.15$& -38 &  3.61 & 1.48&  134  &       20.3    &       99.6    &             &             \\
                 &     &                &     &     &   A3    &4.3$\pm0.2$ & -38 &$<$1.03&     &       &        5.9    &        6.9    &             &             \\  
\object{0153+744}& 4.8 &     1.04       &  10 &  39 &   A     &     0      &  0  &   1.50& 0.60&  119  &      196.4    &      348.8    &             &    0.29     \\
                 &     &                &     &     &   B     &10.1$\pm0.3$& 156 &   1.68& 0.65&  118  &      270.8    &      530.2    &             &    1.70     \\
                 &     &                &     &     &   C     & 7.4$\pm0.4$&     &   3.43& 1.10&  154  &       20.4    &       85.4    &             &    2.04     \\
                 &     &                &     &     &   D     & 4.0$\pm0.4$&     &   1.40& 0.95&  159  &        8.9    &       11.6    &             &             \\
                 &     &                &     &     &   E     & 9.3$\pm0.4$&     &$<$1.00&     &       &       46.4    &       54.0    &             &    2.16     \\
                 & 15  &     0.36       &  10 &  39 &   A1    &     0      &  0  &$<$0.19&     &       &      186.7    &      195.3    &             &             \\
                 &     &                &     &     &   A2    &0.61$\pm0.1$&  69 &   1.00& 0.27&   98  &       36.4    &       73.5    &             &             \\
                 &     &                &     &     &   A3    &1.21$\pm0.1$&  81 &   0.60& 0.20&  139  &        5.9    &       26.4    &             &             \\
    	         &     &                &     &     &   B     &            &  81 &   1.72& 0.85&  122  &       15.3    &       77.8    &             &             \\
                 &     &                &     &     &   C     &            &     &   2.40& 1.17&  168  &        1.2    &        7.8    &             &             \\
                 &     &                &     &     &   E     &            &     &   1.35& 0.99&   36  &        1.3    &        6.0    &             &             \\

\object{0202+149}& 8.4 &     2.07    &$\sim5$ &  16 &   A     &     0      &  0  &   0.82& 0.23&  124  &     1648.2    &     1831.1    &             &   0.32      \\
                 &     &                &     &     &   D     &4.8$\pm0.2$ & -50 &   2.81& 2.12&   20  &       88.8    &      254.1    &             &   0.58      \\
                 & 15  &     1.63    &$\sim5$ &  16 &   A     &            &     &   0.75& 0.15&  132  &     1151.7    &     1477.1    &             &             \\
                 &     &                &     &     &   D1    &4.9$\pm0.15$& -60 &   1.04& 0.42&   88  &       24.7    &       53.0    &             &             \\
                 &     &                &     &     &   D2    &5.1$\pm0.15$& -45 &   1.83& 0.92&   61  &       20.4    &       82.8    &             &             \\
\object{0859+470}& 15  &     0.64    &$\sim5$ &  21 &   A1    &     0      &  0  &$<$0.74&     &       &      388.4    &      404.3    &             &             \\
                 &     &                &     &     &   A2    &0.56$\pm0.15$&-15 &$<$1.81&     &       &       92.1    &      107.9    &             &             \\
                 &     &                &     &     &   A3    &1.98$\pm0.2$& -10 &   2.07& 0.67&  169  &       15.7    &       63.4    &             &             \\
                 &     &                &     &     &   B     &4.36$\pm0.4$&   4 &   1.78& 1.11&  137  &       13.3    &       67.2    &             &             \\

\object{0945+664}& 1.7 &     0.44       &1500 & 6370&  A$^{*}$&            &     &    119&  49 &       &       20.9    &      316.4    &             &             \\
                 &     &                &     &     &  B$^{*}$&            &     &     73&  63 &       &        4.6    &      126.1    &             &             \\
                 & 4.8 &     0.26       &     & 510 &  A$^{*}$&            &     &     94&  78 &       &       77.2    &      255.4    &             &             \\

\object{0954+556}& 1.7 &     1.82    &$\sim80$& 340 &  A$^{*}$&            &     &    38 &  26 &       &      260.7    &      975.3    &    0.42     &   0.88      \\
                 &     &                &     &     &  B$^{*}$&            &     &    62 &  54 &       &       69.8    &      842.5    &    0.49     &   1.62      \\
                 & 4.8 &     0.87    &$\sim45$& 190 &  A$^{*}$&            &     &    28 &  11 &       &       63.6    &      531.9    &             &             \\
                 &     &                &     &     &  B$^{*}$&            &     &    37 &  20 &       &        8.0    &      338.8    &             &             \\
                 & 15  &     0.21    &$\sim45$& 190 &  A$^{*}$&            &     &    15 &   4 &       &       10.6    &      184.4    &             &             \\
 \hline 
 \end{tabular}
\label{mydata}
\end{table*}

\normalsize

According to the information available in the literature, 
each source was observed at one or more of the following
frequencies -- 1.7~GHz, 4.8~GHz, 8.4~GHz and 15~GHz -- with a recording
band-width of 32~MHz at 64~Mbps, for about 2--3 hours divided into scans
of 44 minutes. 
The shortest baseline is Pie Town -- Los Alamos and 
this may have caused some flux density losses in components with angular 
size $\gtsim 20$~mas at 15~GHz or $\sim 200$~mas at 1.7~GHz. 

The data were correlated with the NRAO VLBA processor at Socorro and reduced 
using AIPS (Astronomical Image Processing System). System temperatures 
measured at the antennas during the observations and gain-elevation tables 
provided by NRAO were used for amplitude calibrations. Amplitudes for 
calibrator sources (\object{3C84}, \object{DA193}) were found to be generally consistent within 3\% 
of expectations.
Data were edited with standard procedures. Global fringe-fitting was 
performed using the AIPS task FRING with solution intervals of about 4 
minutes at 1.7--4.8--8.4~GHz and 1 minute at 15~GHz. All sources provided 
fringes with high signal-to-noise ratio on all baselines except  
\object{0945+664} for which the task FRING could not find solutions with signal 
to noise ratio $\geq 5$ at any of the three frequencies for any 
solution interval. For this source we successfully applied the delay and rate solutions 
found for \object{0954+556} ($\sim 11\degr$ apart from \object{0945+664}), observed in adjacent scans.

Images were obtained with the AIPS task IMAGR after a number of phase 
self-calibration iterations using a 
decreasing solution interval, starting with a few minutes (6 or 4)
down to 0.5 minutes, ending with
a final step of amplitude self-calibration with a solution interval 
longer than the scan length (44 min) in order to remove residual systematic 
errors. This latter step was applied only if the gain corrections were 
$<3$\%. 

\smallskip

The source total flux densities  were 
obtained by integration over the source images by means of AIPS task 
TVSTAT. 
For moderately compact components, Gaussian fits were performed  with the 
AIPS task JMFIT, which provides positions, angular sizes and total flux
density. Occasionally we made use of images at moderate super-resolution. 
Component positions are given with respect to the brightest component 
in the image, in order to allow a comparison with similar
data at other epochs and search for proper motions. As JMFIT gives errors 
on the various parameters which tend to be unrealistically small,  
we did not use them. Instead we estimated empirically the parameter 
uncertainties from variations of the parameters during the model fitting 
process. 

For many resolved sources no positions are given and the angular sizes are 
taken from the lowest reliable contours in the images. 

Images for spectral index information (not shown) were obtained at the 
different
frequencies using the \textit{uv}-range between the shortest baseline at
high frequency and the longest baseline at low frequency and convolving
to the lowest resolution. Alignment of the images was obtained by using a
compact feature as a reference.
Using the package SYNAGE++ (Murgia \& Fanti \cite{Murgia96}), which
allows one to interactively select
different source regions on the images, we obtained {\it local spectral
index} values between pairs of frequencies ($\alpha_{lo}$ and $\alpha_{hi}$
between low and high frequency pairs respectively, in Table ~\ref{mydata}). 
 
 Images of total intensity are shown in Fig. ~\ref{images}  and source parameters 
 are given in Table ~\ref{mydata}.

\section{Comments on individual sources}

{\bf \object{0133+476}}. The source integrated spectrum  (K\udier hr et al. \cite{Kuhr81}) is flat
with a 20-30\% variability at frequencies higher than 1~GHz. From
earlier VLBI observations (Pearson \& Readhead \cite{PR81}, Polatidis et al. \cite{Polatidis95}) 
there were indications  for classifying the source as a \emph{core-jet}.
At 15~GHz (Kellermann et al. \cite{Kellermann98}) the source shows a very bright and compact
component plus a fainter secondary one at a distance $\approx$ 3~mas in P.A.$\sim-45\degr$, 
with a very large flux density ratio and some 
weaker emission in between. Although the authors classified it as a
\emph{single sided source}, due to its similarity with OQ208 which is a
CSO with very unequal flux density components, we decided to investigate it further
at several frequencies.

Our images at 1.7, 4.8 and 8.4~GHz show without any doubt that the structure
is indeed of the \emph{core-jet} type.
The  radio emission is dominated by a bright compact component 
(component $A1$ at 8.4~GHz) and a jet in P.A.$\sim -45\degr$ which 
diffuses, as seen often in many core-jet sources, into a sort of tail with several wiggles, 
seen at 5~GHz and 1.7~GHz North 
of the source.

\begin{figure}
\begin{center}
  \includegraphics{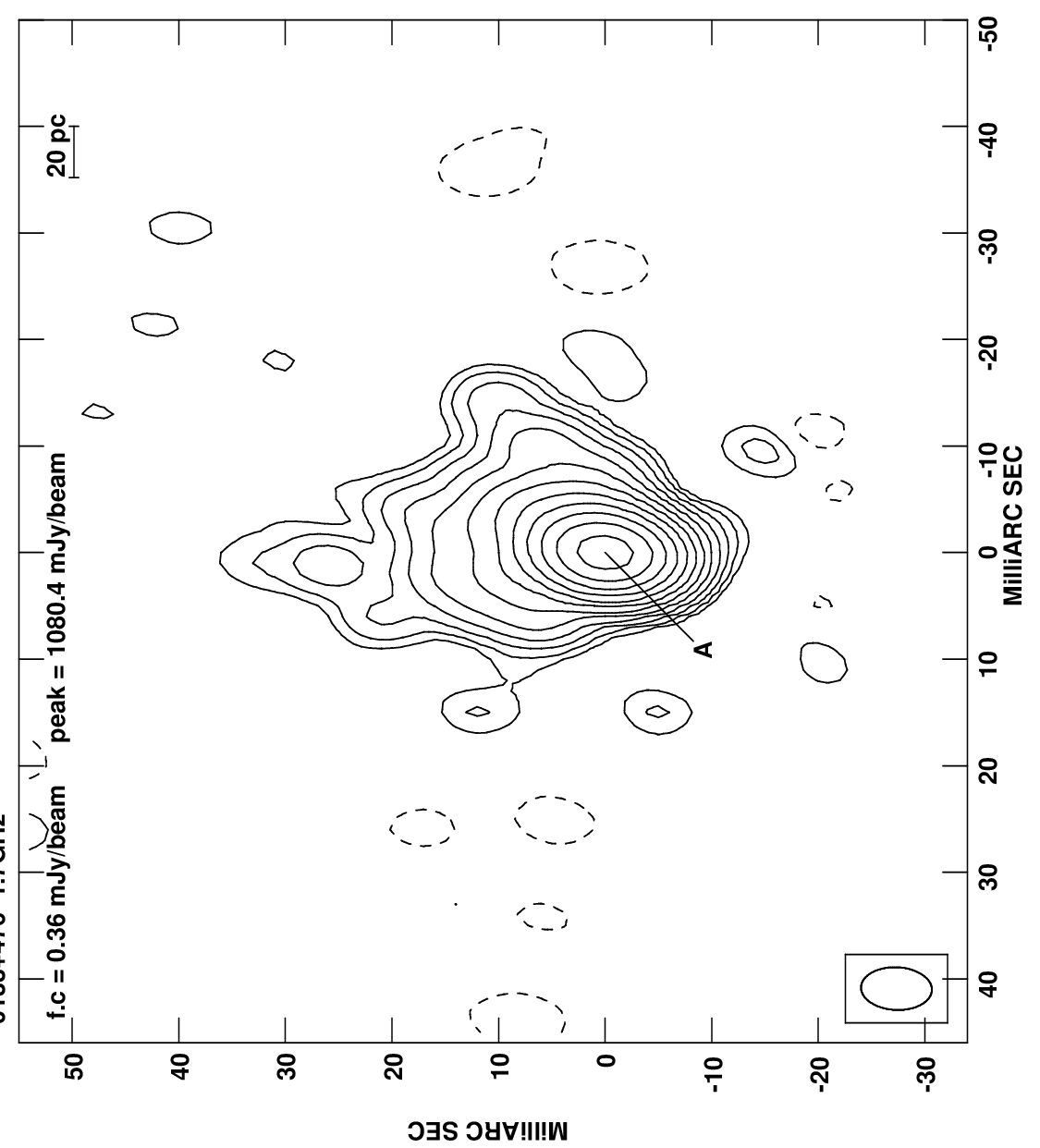}
  \includegraphics{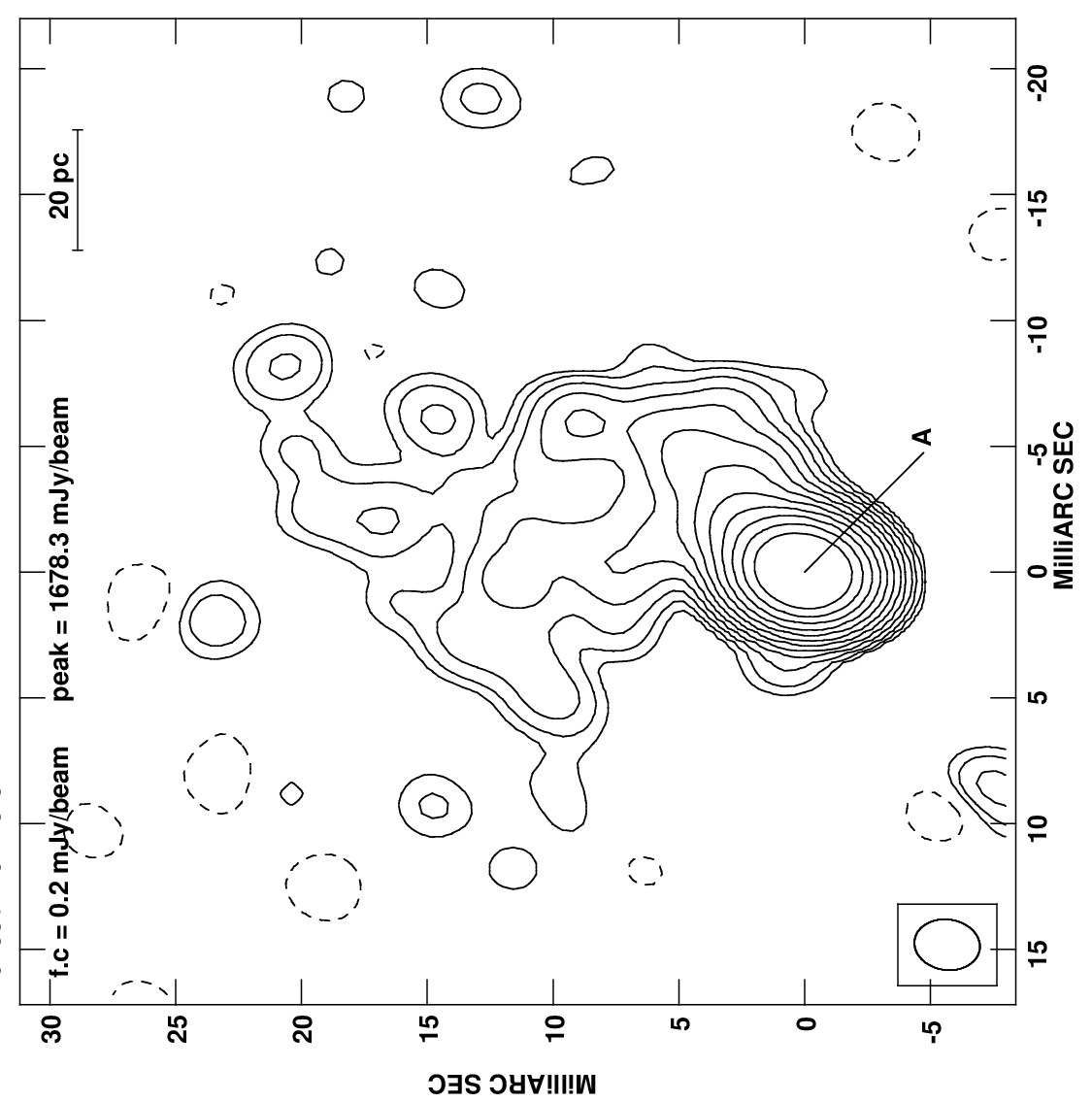}
  \includegraphics{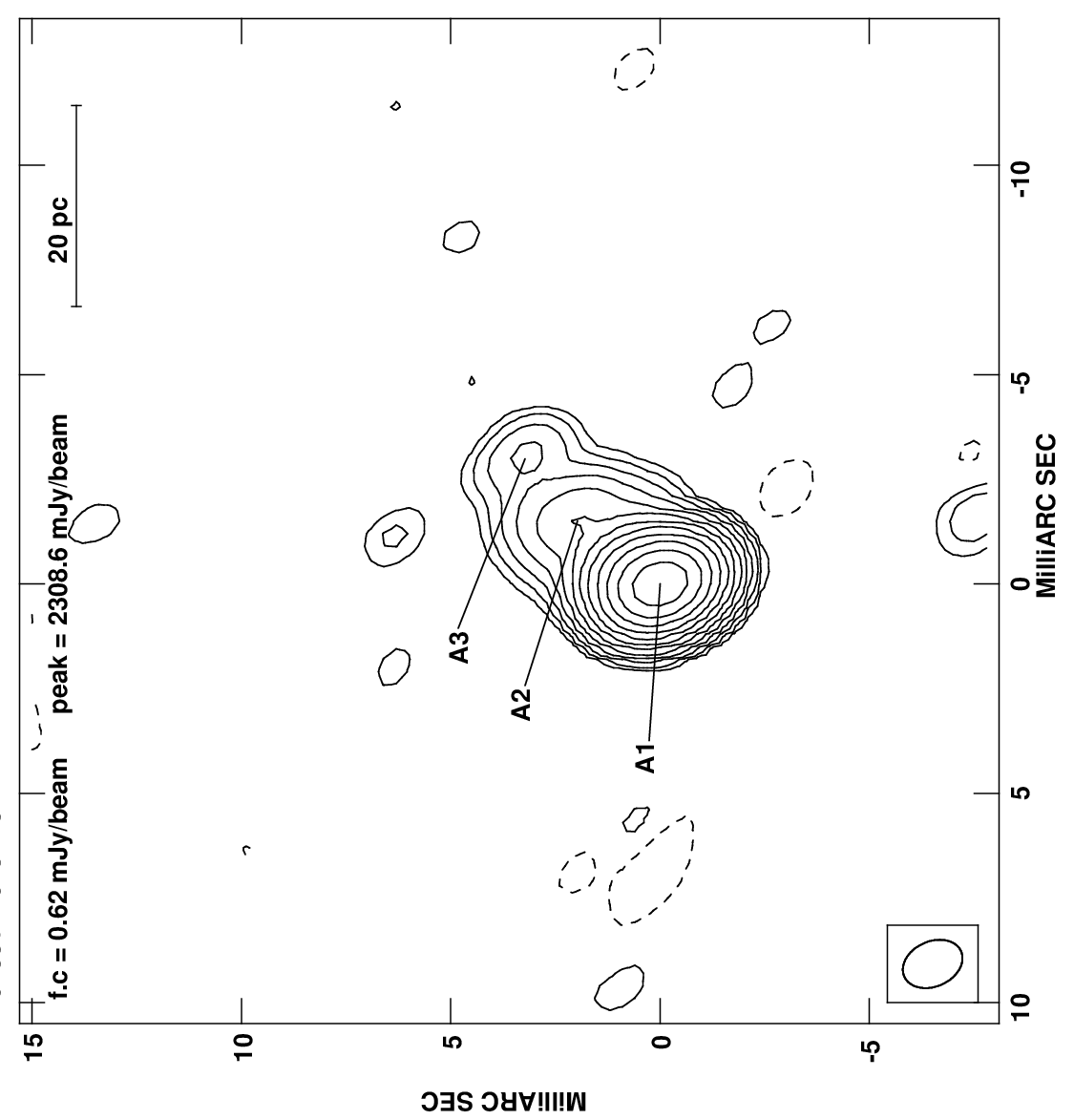}
  \vspace{22.6cm}
\caption{VLBA images: the first contour (f.c.) is generally three times the rms noise level on the image; contour levels increase by a factor of 2; the restoring beam is shown in the bottom left corner of each image.}
\label{images}
\end{center}
\end{figure}

 The images, made with the same \textit{uv}-range and restoring beam as the
 1.7~GHz observation, allowed us
 to obtain the spectral index of component $A$, which is 
 only marginally resolved at 1.7~GHz. We
 find $\alpha_{1.7}^{4.8}=-0.37$, $\alpha_{4.8}^{8.4}=-0.55$, the 
 spectral index being less inverted at low frequency probably because
 of a larger 
 contribution to the total flux density from the jet/tail emission.

We compared the integrated flux densities in our images with those from the 
University of Michigan Radio Astronomy Observatory (UMRAO) database.
Due to the source variability we took data close in
time to our observing period ($\pm 2$~months). 
At 4.8~GHz our flux density is within 3\% of what is expected, 
while at 8.4~GHz $\approx 10\%$ of the total flux density is missing in our 
VLBA image, clearly due to the absence of short \textit{uv}-spacings. 
The integrated spectrum of the source is definitely inverted between 1.7 
and 8.4~GHz, so that the source, at our observing epoch, would be classified as a
\emph{High Frequency Peaker} (Dallacasa \cite{Dallacasa03}).
 
Comparing our 8.4~GHz flux density of component $A1$ with the peak flux 
density of the 
bright component in the 15~GHz map of Kellermann et al. (\cite{Kellermann98}) we get an 
estimate of the spectral index $\alpha_{8.4}^{15} \le 0.3$. This would 
suggest that the spectrum peaks somewhere in between these two frequencies. 
However we should consider that the two images were taken about two years 
apart so that the result may be affected by variability.

Component $A2$ is likely to be the faint component in the 15~GHz image
of Kellermann et al. (\cite{Kellermann04}), as the separations from the brightest component are
similar in the two images. Kellermann et al. (\cite{Kellermann04}) analyse the separation between $A1$
and $A2$ at 15~GHz, from 1994.5 to 1999.0, and give a proper motion of
0.04$\pm$0.01~mas/yr ($\beta_{app}=1.01\pm0.25$ with the cosmological parameters we use). 
The separation we measure (2.5$\pm$0.2~mas) is lower than
the extrapolation of their proper motion at our epoch at a 1.8 sigma level.
This difference, although not significant, may be due to spectral 
effects within $A1$ (see e.g. Marcaide \& Shapiro \cite{Marcaide84}), as our image is at a frequency lower than theirs.
Alternatively it may indicate that the true proper motion may be smaller than the one they reported.

\bigskip

{\bf \object{0153+744}}. This source was studied earlier by Hummel et al. (\cite{Hummel88}, 
\cite{Hummel97}), at several frequencies. Images at 2.4 and 8.5~GHz can be found in 
Fey et al. (\cite{Fey97}). 
The source integrated spectrum (K\udier hr et al. \cite{Kuhr81})  is
flattish ($\alpha \approx 0.3$) with some evidence of variability.
The radio emission  at cm 
wavelengths is dominated by two main components, $A$ and $B$ with a knotty  
 ``bridge'' of emission (components $C$, $D$, $E$ as labelled in Hummel et al. \cite{Hummel97}). 
At low resolution and lower dynamic range the source resembles a small double with components of
similar flux densities.
In our 15~GHz image component $A$ is resolved into three components plus a 
diffuse and faint outer emission region,
with the brightest component at the western edge, giving the appearance of 
a core-jet structure, in agreement with what found by Hummel et al. (\cite{Hummel97})
at 22~GHz. 

Comparing our 15~GHz map with that of Hummel et al. (\cite{Hummel97}) at 22~GHz we
find no significant motion for component $A2$, while component
$A3$ may have moved away from $A1$ by 0.2$\pm 0.1$~mas over 6.6 years,
leading to a proper motion of $0.035\pm 0.017$~mas/yr and an apparent
velocity $\beta_{app} = 1.5 \pm 0.7$.
     	   
From the comparison of our 4.8~GHz measurements with those reported by Hummel et al.
 (\cite{Hummel88}, \cite {Hummel97}), we find that component $A$ dropped by 
about a factor of two in flux density after 1990. It is likely 
that similar drops have occurred at other wavelengths as well so that it 
is not possible to determine a reliable spectral index of component $A$  over 
a broad frequency range from literature data taken at different epochs. 
From our own nearly simultaneous observations a spectral index $\alpha^{15}_{4.8} 
\approx 0.29$ is derived at the resolution of the 4.8~GHz image.
We find a spectral index $\alpha^{15}_{4.8}=-0.19$ at the position of
component $A1$ which steepens away toward the East of component $A$ up
to $\alpha^{15}_{4.8} \approx 1.2$.

\begin{figure}[htbp]
\addtocounter{figure}{-1}
\begin{center}
  \includegraphics{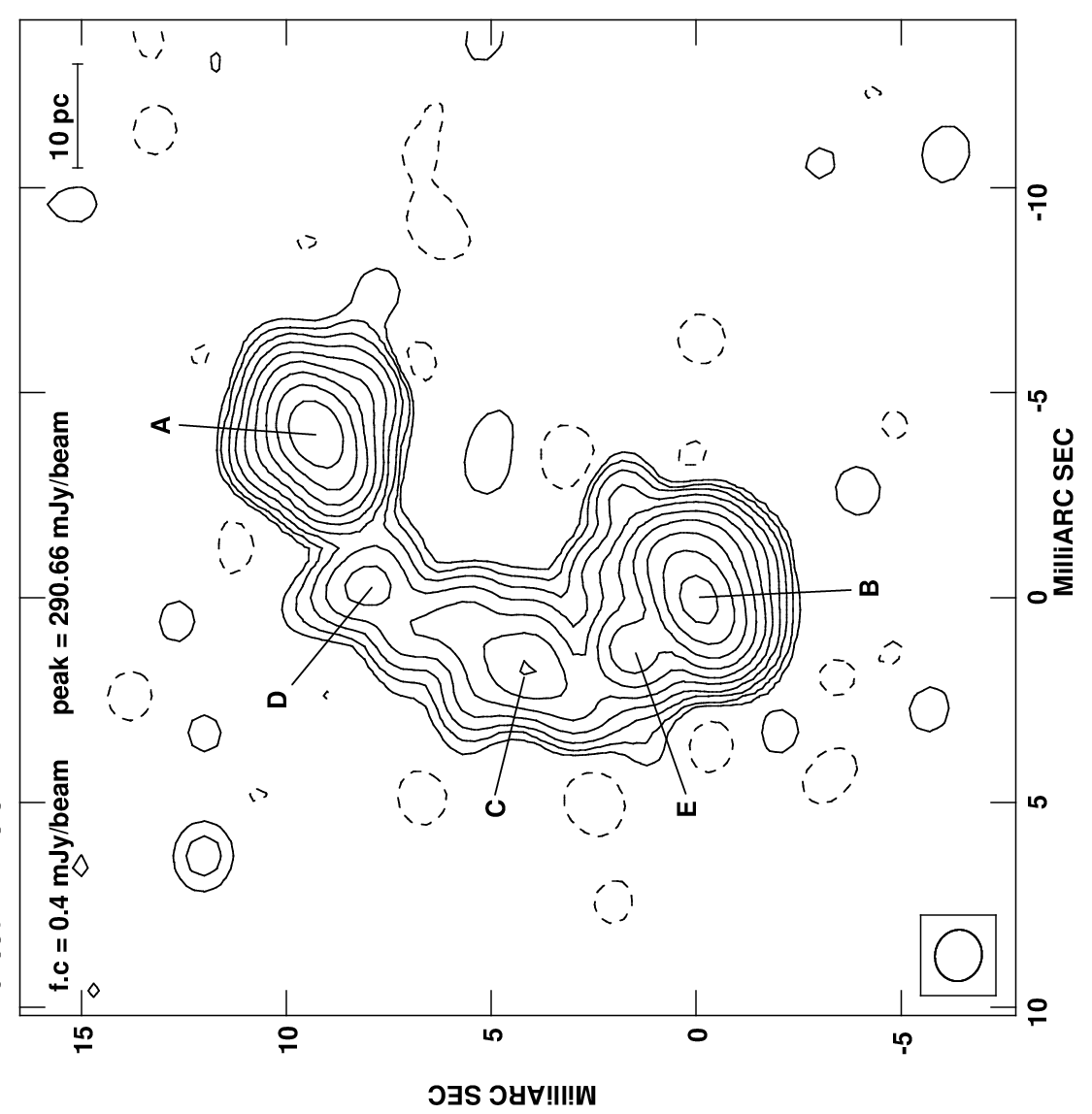}
  \includegraphics{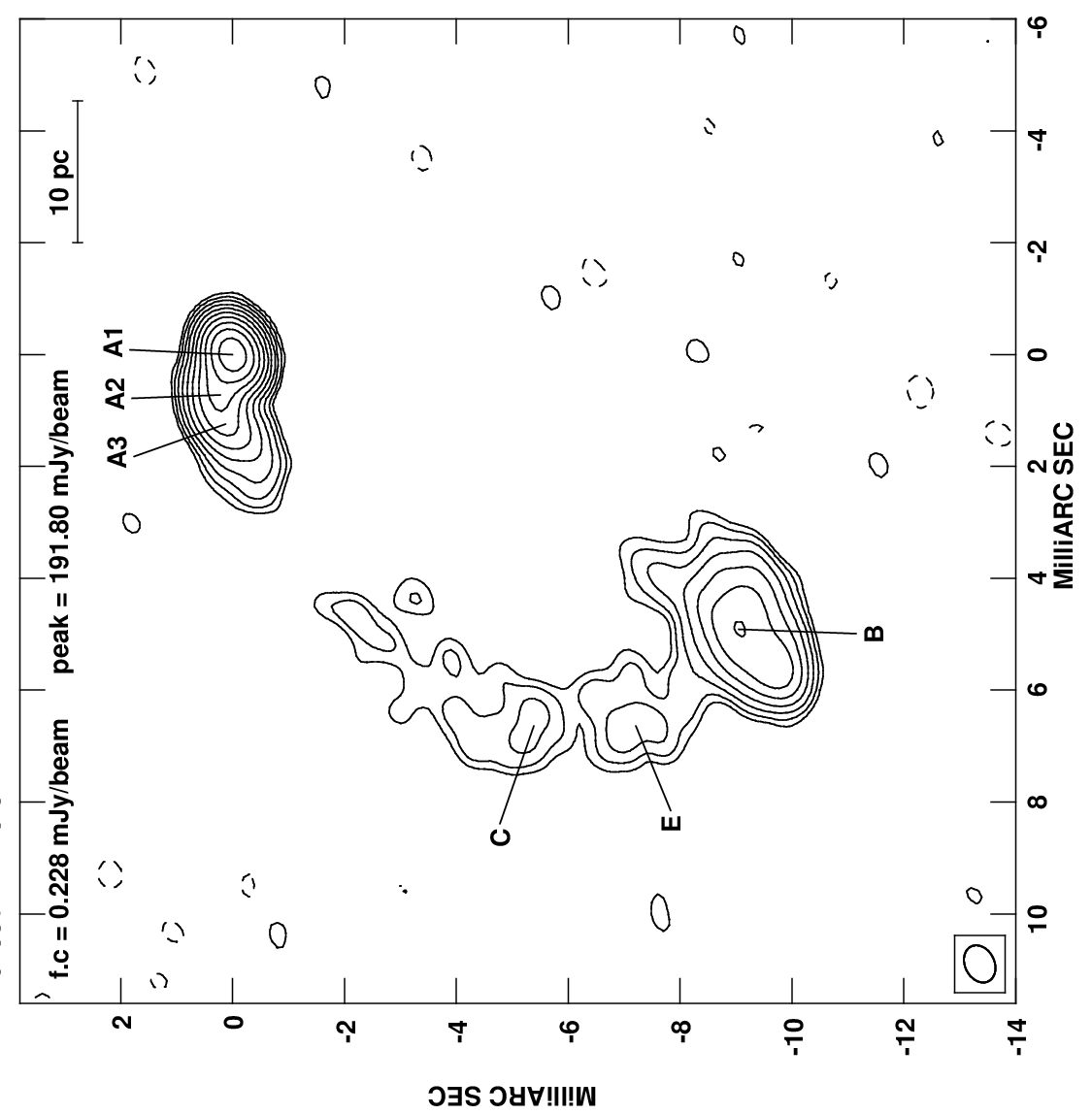}  
\vspace {15.0cm}
\caption{continued.}
\end{center}
\end{figure}

Component $B$ shows a complex 
 structure consisting of resolved steep-spectrum emission, embedded in a more diffuse component.
 From a comparison with Hummel et al. (\cite{Hummel97}) there is no evidence
 for flux density variability.
 From our flux densities we get a spectral index
 $\alpha_{4.8}^{15} \approx 1.7$. The data of Hummel et al. (\cite{Hummel88}) at 1.7~GHz
 and of Fey et al. (\cite{Fey97}) at 2.3~GHz,  
 taken at lower  resolutions, imply that the spectrum 
 flattens significantly at lower frequencies ($\alpha^{4.8}_{1.7} \approx 
 0.5$).
 Also, components $C$ and $E$ seem to have a steep high frequency spectrum. Hummel et al. 
(\cite{Hummel97}) have examined the separation of components $A$ and $B$
at several epochs and concluded that it has not changed over a 10 year
interval. Our data confirm that the separation is still unchanged over 20
years  giving a proper motion of 0.008$\pm$0.007~ mas/yr, 
corresponding to $\beta_{app}=0.36\pm0.31$.

\begin{figure*}[htbp]
\addtocounter{figure}{-1}
\begin{center}
  \includegraphics{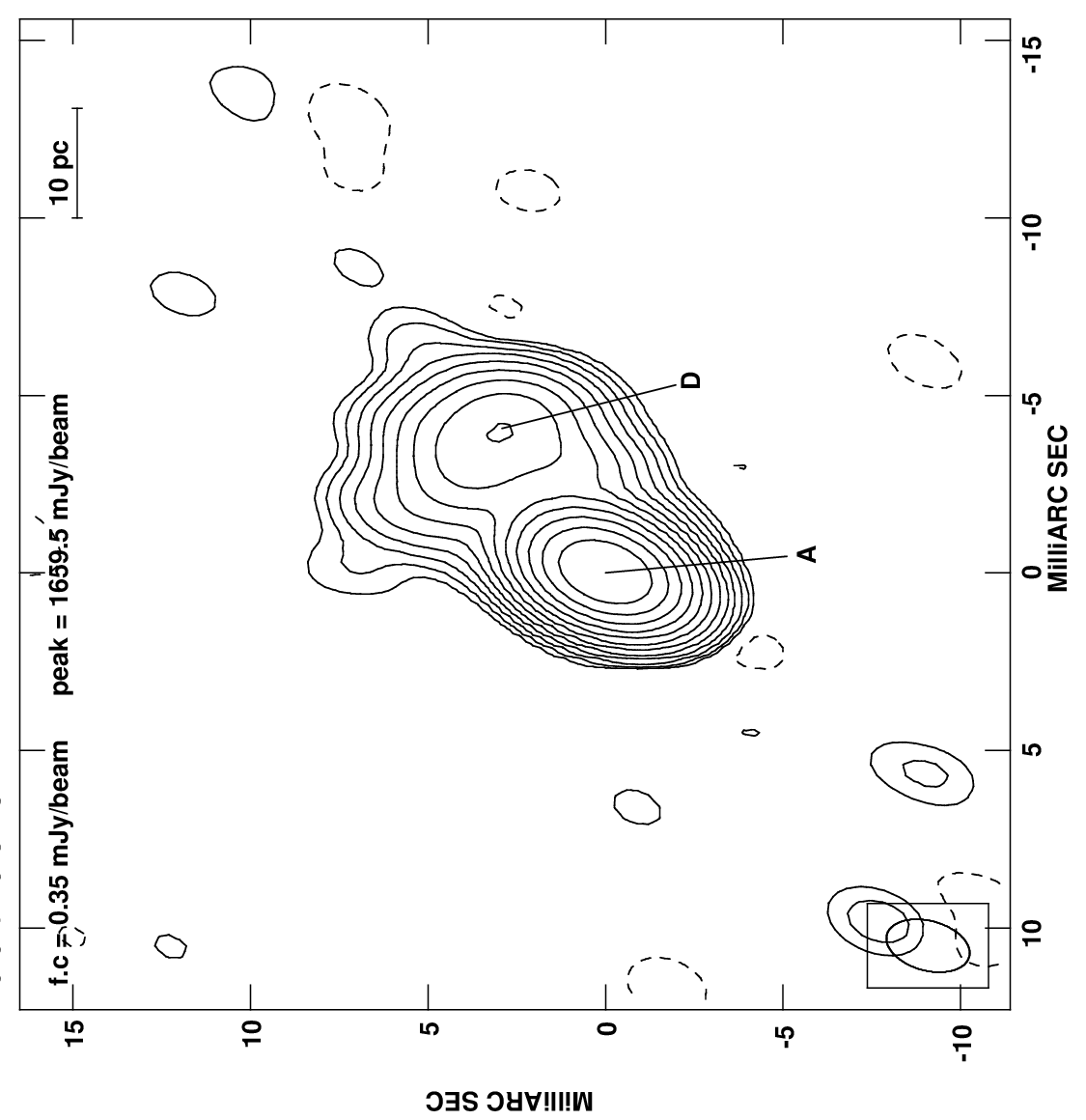}
  \includegraphics{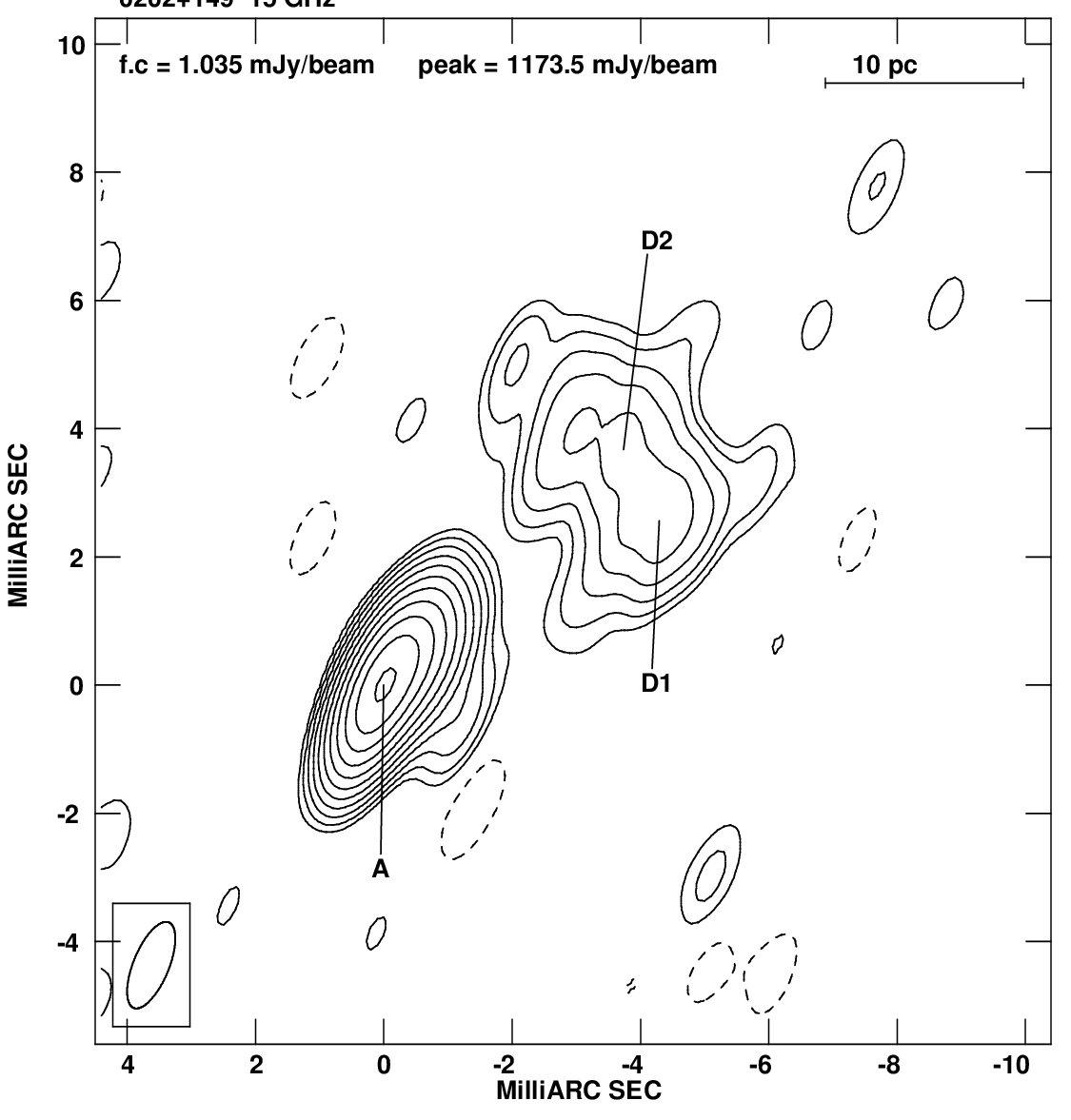}

\vspace {7.5cm}
\caption{continued.}
\end{center}
\end{figure*}

\bigskip

{\bf \object{0202+149}}. There is some confusion in the literature about the
classification of the optical counterpart and redshift.  
By detecting two emission features in the optical spectrum, this object has 
been classified as quasar, with z=0.833 (Stickel et al. \cite{Stickel96}).  
Comastri et al. (\cite{Comastri97}) report a redshift of 1.202.
The most recent information is from Perlman et al. (\cite{Perlman98}), who published an 
optical spectrum which shows several narrow emission lines 
(e.g., H$\alpha$, H$\beta$, [OIII] Mg II), typical of low ionization narrow 
line radio galaxies and derived z=0.405. We take this as the most reliable 
red-shift.
Stickel et al. (\cite{Stickel96}) and Impey \& Tapia (\cite{TI90}) report respectively red and visual
magnitudes $m_{\mathrm R} \approx$ 21.3 and $m_{\mathrm v} \approx$ 20.9. 
Absolute magnitudes would be $M_{\mathrm R} \approx$ -20.4 and 
$M_{\mathrm v} \approx$ = -20.8, rather faint for a quasar.
Fugmann \& Meisenheimer (\cite{FM88}) describe the optical spectrum as a power law with 
spectral index $\alpha \approx 2.3$. They also show that the infrared spectrum is consistent
with the same spectral index, although the source is very variable at these
wavelengths.
The source is identified with a $\gamma$-ray source (von Montigny et
al. \cite{Montigny95}).

The source has a flattish radio spectrum ($\alpha \approx 0.3$)
with some variability across the whole radio band. 

In the image at 15~GHz by Kellermann et al. (\cite{Kellermann98}) the source shows a very
bright compact component ($A$) plus a diffuse component ($D$) at $\approx$
5 mas distance.
In between the two, at P.A. $\approx -80\degr$ and closer to the bright 
component, another fainter and compact component is seen.
Although the authors described the source as a core-jet  we thought that it
could be an asymmetric double with a weak core and therefore worth 
observing again. 
Additional images at 2.3 and 8.5~GHz from geodetic VLBI observations
(Piner \& Kingham \cite{Piner98}, Pyatunina et al. \cite{ Pyatunina00}) and from VLBA observations at 15,
 22~GHz (Wiik et al. \cite{Wiik01}) and 43 GHz (Pyatunina et al. \cite{Pyatunina00}) are also  available.
From these images our component $A$ turned out to have a rather complex structure,
being resolved into a bright component (called $A$ in Pyatunina et al. \cite{Pyatunina00})
plus two weaker ones
($B$ and $C$ in Pyatunina et al. \cite{Pyatunina00}),
which show activity and possibly superluminal motions.
 
Our images at 8.4 and 15~GHz are consistent with the other images as far as
the broad structure is concerned. The very weak south-western extension of $A$ 
could be related to the faint intermediate component in Kellermann et al. (\cite{Kellermann98})
and component  $C$ in Pyatunina et al. (\cite{Pyatunina00}). 
Compared with the 2~GHz data of Piner et al. (\cite{Piner98}), with a resolution
similar to ours, the separation we measure at 8.4~GHz between $A$ and $D$
strengthens its constancy over a time baseline of ten years, giving a proper motion 
of 0.02$\pm$0.03~mas/yr, corresponding to $\beta_{app}\approx 0.3\pm 0.44$.
Furthermore, our 15~GHz image shows a separation between $A$ and the bright features
$D1$ and $D2$ equal to that reported by Kellermann et al. (\cite{Kellermann04})
at the same frequency over a six-year interval, confirming that no apparent motion is present.
At our epoch of observation, components $A$ and $D$ had spectral indices 
$\alpha_{8.4}^{15}$ $\approx$0.32 and $\approx$ 0.58 respectively.

In our 15~GHz image, component $A$ is slightly extended toward $D$.
A two-component fit gives a bright component with a flux density $\approx 1200$~mJy that we call $A1$
and a fainter component (270~mJy at $0.65\pm 0.20$~mas from $A1$ in P.A.$\sim -53\degr$) that we call $B$.
This $B$ component could be
the homonymous component for which Pyatunina et al. (\cite{Pyatunina00}) and 
Kellermann et al. (\cite{Kellermann04}) give proper motions of 0.18$\pm$0.01~mas/yr
and 0.25$\pm$0.06~mas/yr respectively, 
and $\beta_{app}=2.7\pm0.15$, $\beta_{app}=3.8\pm0.9$ 
with our adopted cosmological parameters.
The radial distance we measure is slightly smaller but compatible with the extrapolation
from Pyatunina et al. (\cite{Pyatunina00}) to our observing
date. Instead it is smaller at the 2.8 $\sigma$
level compared with the extrapolation from Kellermann et al. (\cite{Kellermann04}).
This may point to a lower expansion rate. However, we note that there are 
differences of up to 20$\degr$ in the reported position angles, which may cast some doubt 
on the component identification or mark a rather complex situation about the source 
structure and/or motions.
However, in Kellermann et al. (\cite{Kellermann04}) component $B$ has a low 
rating in their quality marks.
 
We note that in Piner \& Kingham (\cite{Piner98}) and in Pyatunina et al. (\cite{Pyatunina00}) 
components $A1$  and $B$ both  seem to have a flat spectrum 
between 8.4 and 15~GHz. Instead in the range 8.4 -- 43~GHz the brighter
$A1$ seems to have a transparent spectrum, while $B$ 
seems to be still flat. 
Although no errors are quoted for the flux densities and therefore the
spectral indices may be uncertain, this 
suggests that one should not exclude the alternative interpretation that the core is 
component $B$ (called $C_1$ in Piner \& Kingham \cite{Piner98}) and therefore that the source 
could be a very asymmetric CSO. The asymmetries in component size 
and flux density (11:1) are not dissimilar to those seen in other 
very small, tens of pc, two-sided sources.

\bigskip

{\bf \object{0859+470}}. In our 15~GHz image this quasar is characterized 
by a bright compact component and a more diffuse tail-like emission 
extending out to $\sim10$ mas to the North.
The radio spectral index is $\alpha\sim 0.3$. As a result 
the object is a ``classical'' flat spectrum quasar.   

\begin{figure}[htbp]
\addtocounter{figure}{-1}
\begin{center}
\includegraphics{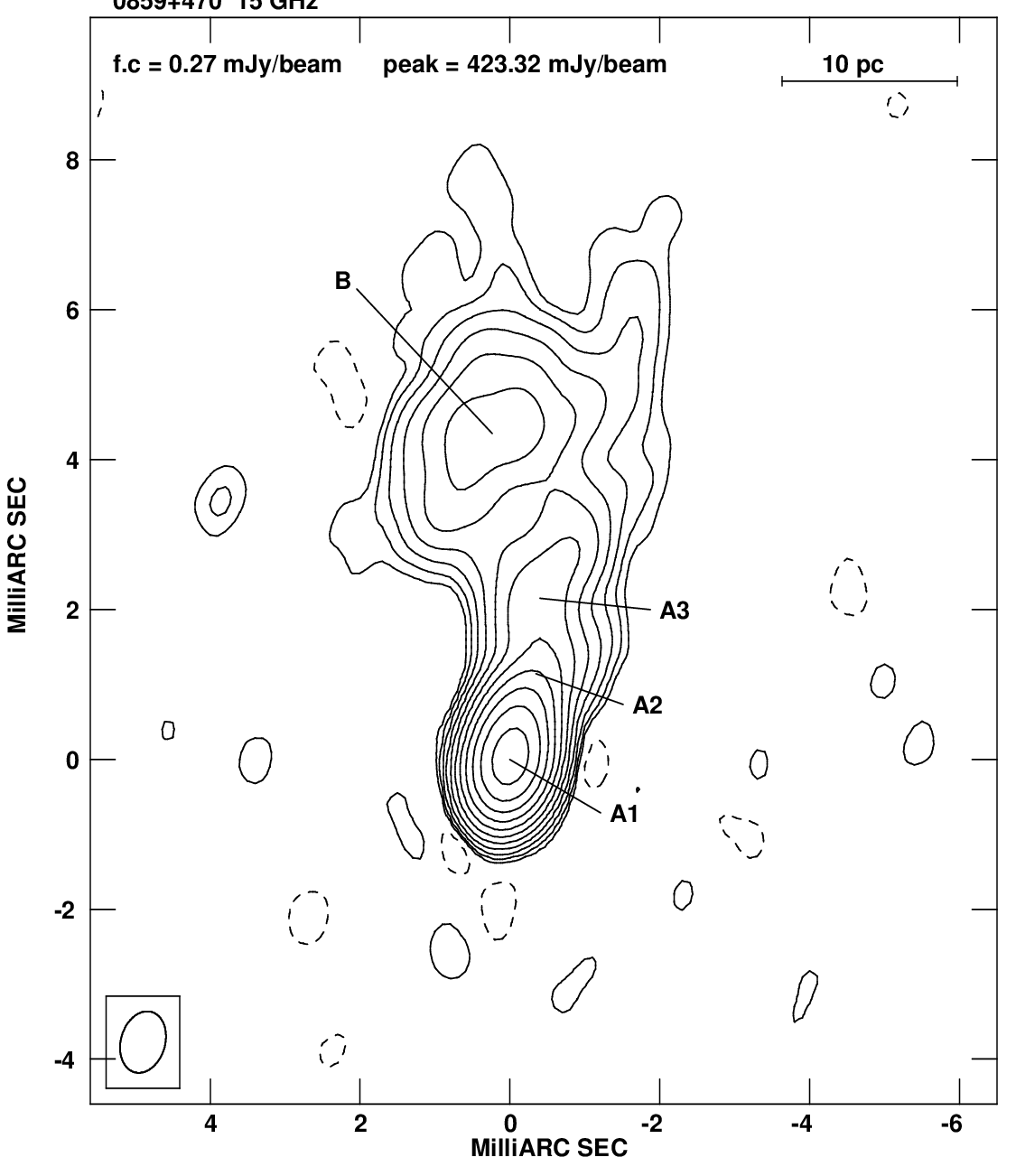}
\vspace{8.0cm}
\caption{continued.}
\end{center}
\end{figure}

We have compared the separations of components $A1$, $A2$, $A3$ and $B$ with
those given in Fey et al. (\cite{Fey97}) at 8.4~GHz, at a resolution similar
to our own. For that purpose we have also convolved our image to their
resolution.
The two observations were carried out $\approx$ 4.6 years apart.
Our component $B$ is very likely component 4 in Fey et al. (\cite{Fey97}), and our component
$A3$ could be their component 2, in which case it would have moved away from $A1$ by
$0.38\pm 0.15$~mas with $\beta_{app}= 2.8\pm 1.1$. 
Their component 3 is not seen in our image and it might
have moved out and now be confused with component $B$.
Component $A2$ might be a new one, not seen in 1995. 
Although two epochs only do not allow us to firmly determine proper motions,
there is no doubt that structure changes have occurred between the two
epochs.

\bigskip

{\bf \object{0945+664}}. The spectral index of the source (K\udier hr et al. \cite{Kuhr81})
is $\approx$ 0.5 at
frequencies below 5~GHz and then it steepens gently to $\approx$ 0.75
above 10~GHz. Xu et al. (\cite{Xu95}) presented a VLA image at 1.4~GHz, with a
resolution of $1\farcs5$, where the source is slightly resolved, possibly double, 
with a separation $\approx 1\farcs5$. 

\begin{figure}[htbp]
\addtocounter{figure}{-1}
\begin{center}
\includegraphics{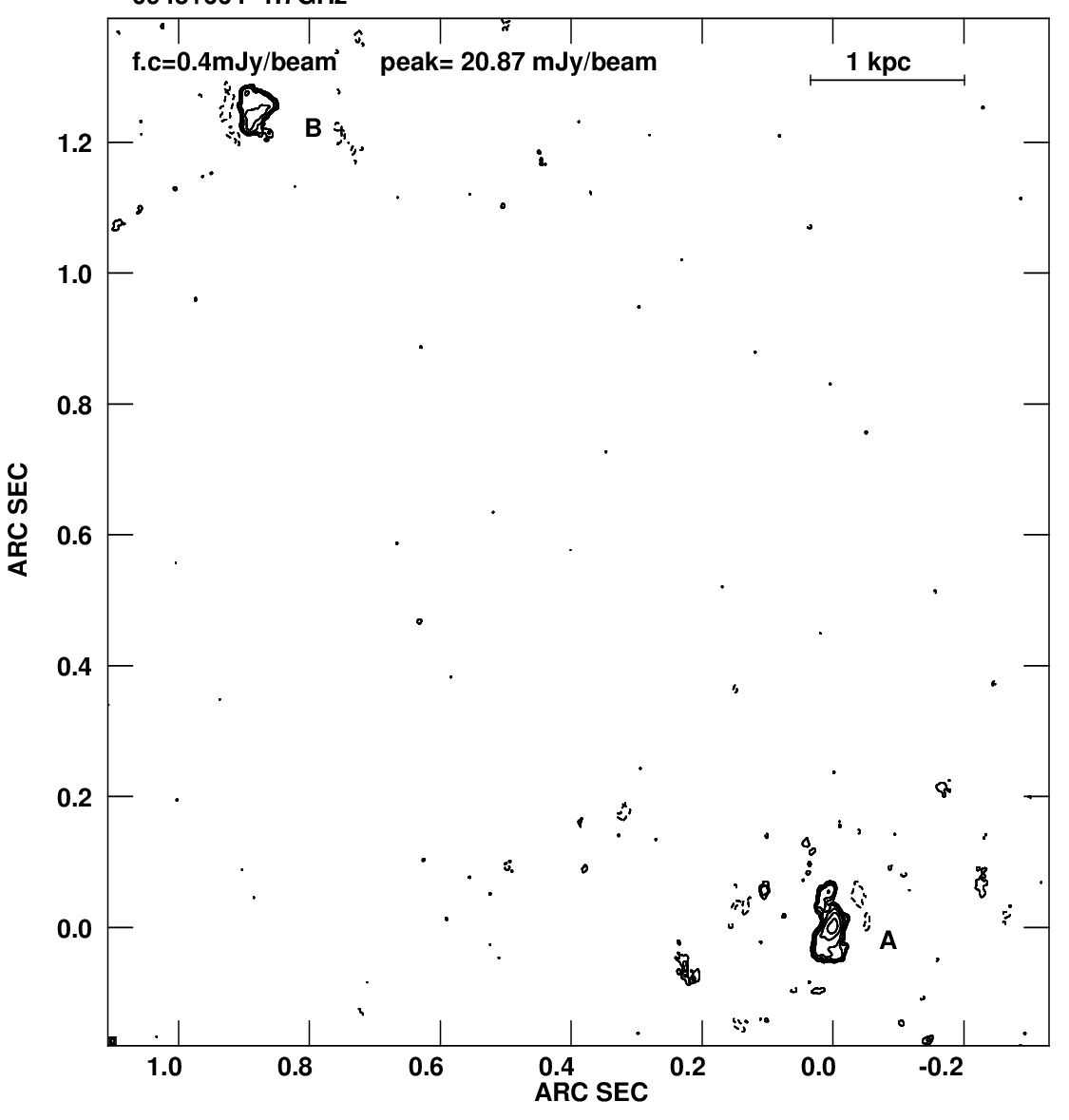}
\includegraphics{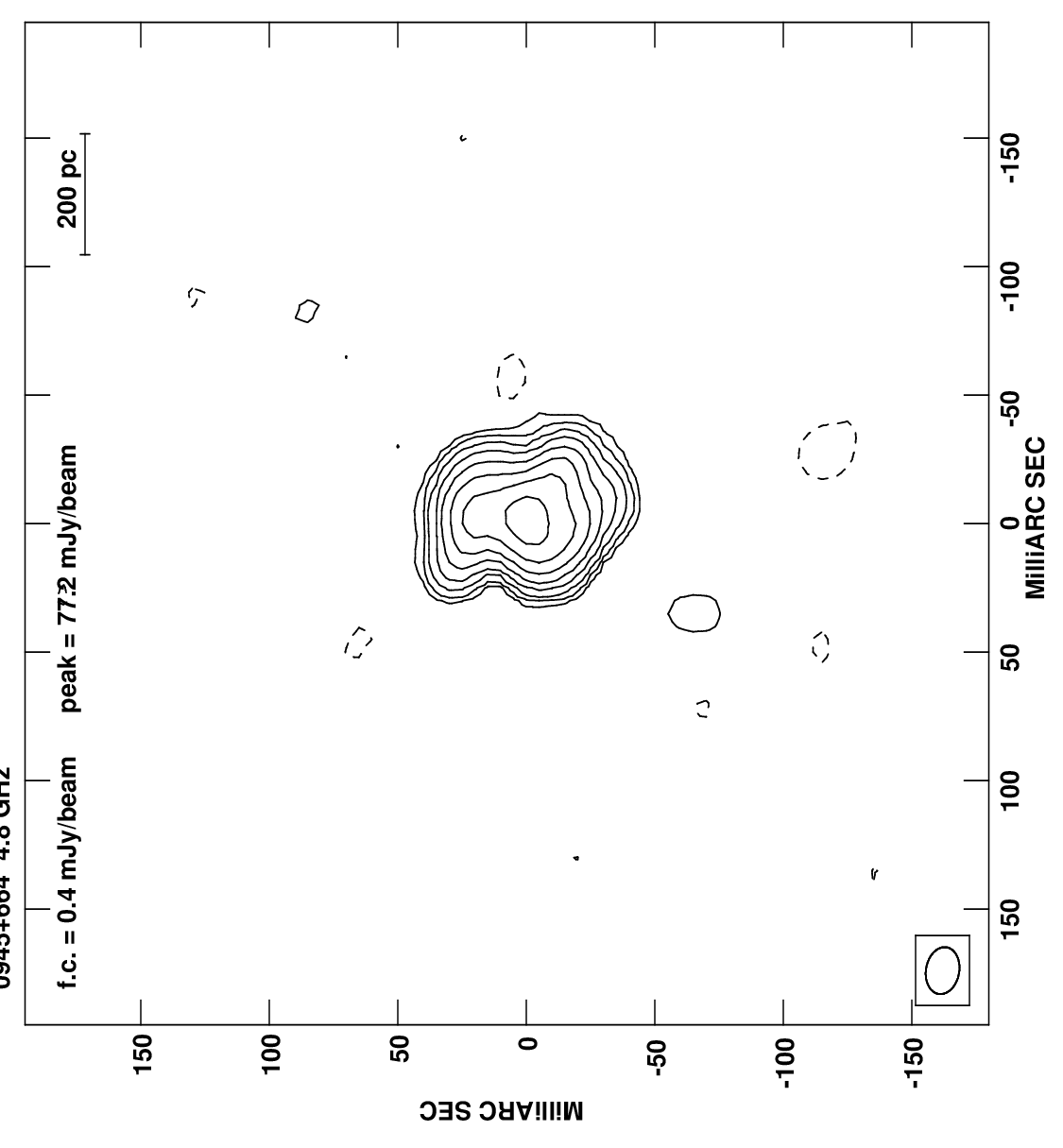}
\vspace{15.5cm}
\caption{continued.}
\end{center}
\end{figure}

Unpublished images at 5~GHz (MERLIN, see NED), 8.4 and 15~GHz (VLA, J. Browne, private communication)
show the double structure better and allow us to determine the spectral indices of the radio lobes
(component $A$ has $\alpha^{15}_{8.4}$=0.8 and component $B$  
$\alpha_{1.7}^{4.8}$=1).

The image we obtained at 1.7~GHz shows two components, $A$ and $B$, separated 
by $\sim 1\farcs5$ in P.A.$\sim 35$\degr, well resolved 
by VLBI observations which are the hot spots of the two lobes. 
This image accounts for  20\% only of the total 
flux density of this source. 

At 4.8~GHz we detected one component only, very likely the one stronger at 1.7~GHz
(component $A$), and on a small subset of short baseline only.
We present in Fig. ~\ref{images} a tapered image, again accounting for 20\% of 
the source total flux density at this frequency.

At 15~GHz also component $A$ is completely resolved out and the image revealed 
no significant radio emission. 
We conclude that the object 
is an MSO, even if the core component has not been identified.

\bigskip

{\bf \object{0954+556}}. The integrated spectrum (K\udier hr et al. \cite{Kuhr81}) is straight up to
30~GHz, with a spectral index $\alpha \approx 0.4$ and no sign of variability.
Our 1.7 and 4.8~GHz images  show radio emission distributed over
two fat components separated by $\sim 50$~mas. 
The North component ($A$) contains a bright compact hot spot, while the South 
component ($B$) is characterized by a more diffuse emission which is resolved 
out at 15~GHz.

The integrated flux densities from our images account
only for 60\%, 43\% and 20\% of the total flux density of the source at
1.7, 4.8 and 15~GHz respectively.
Convolving the two higher frequency images to the same resolution of the one
at 1.7~GHz we have obtained the three-frequency spectrum of the northern compact hot spot.
We find $\alpha \approx 0.4$.   

\begin{figure}[htbp]
\addtocounter{figure}{-1}
\begin{center}
\includegraphics{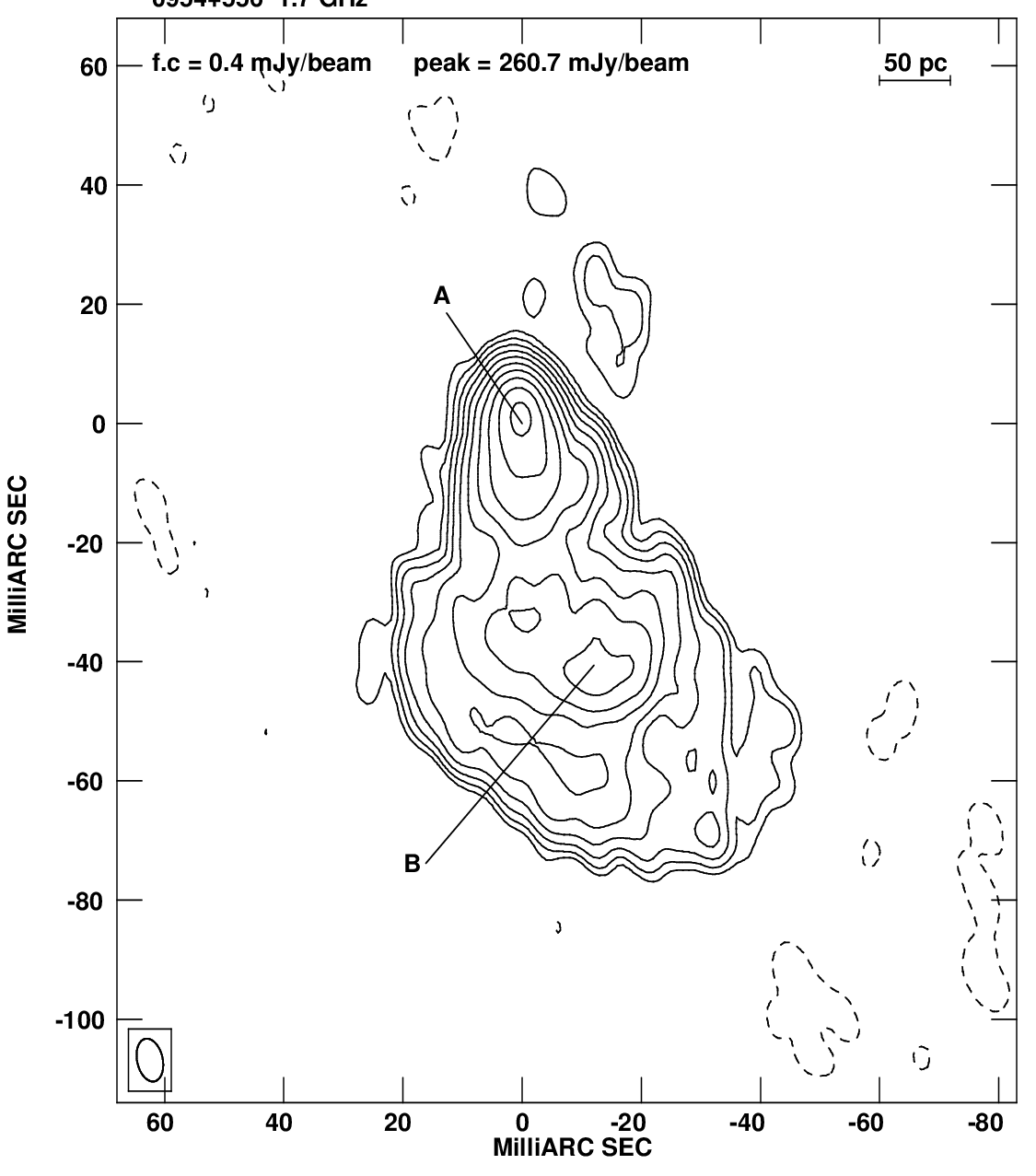}
\includegraphics{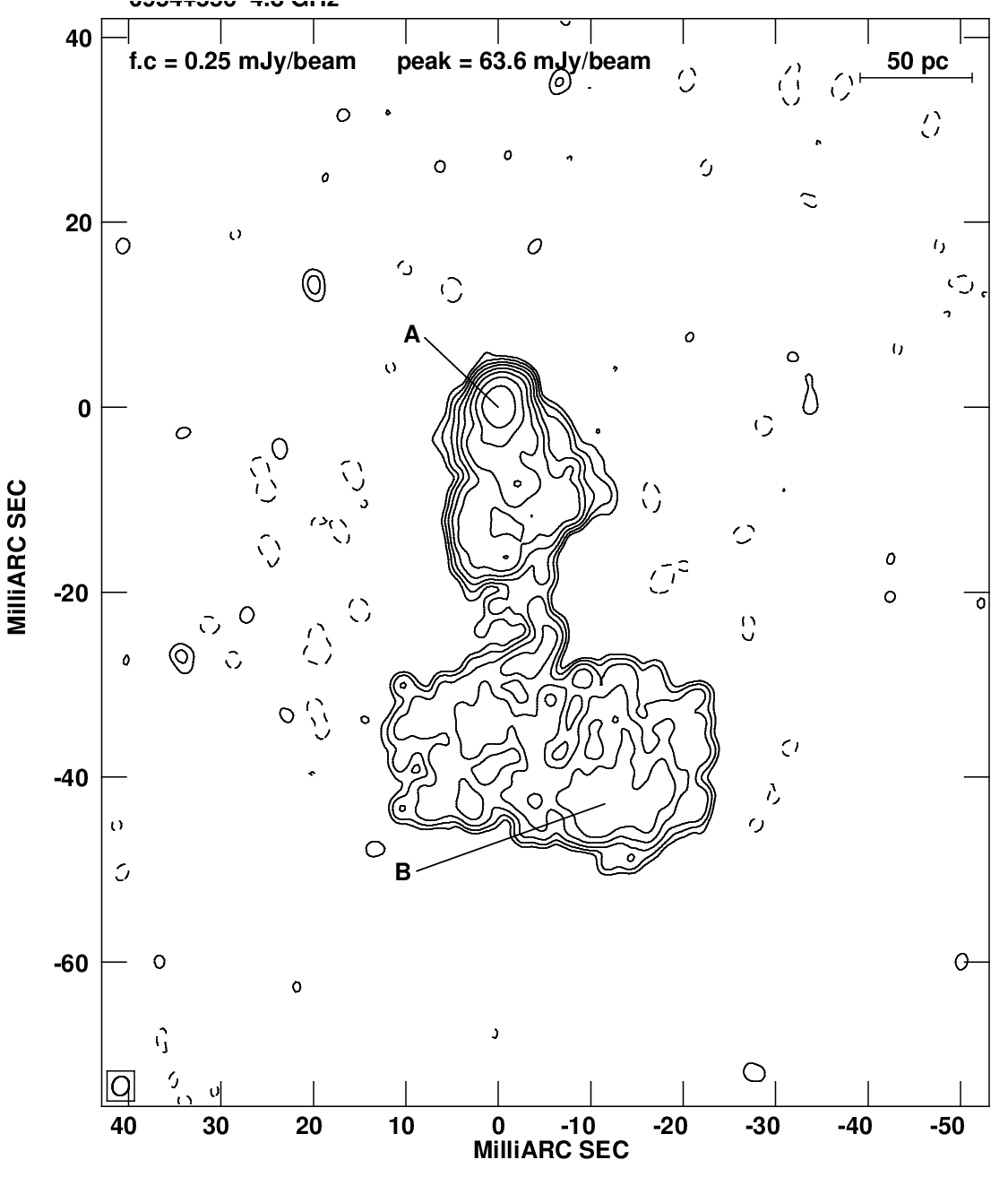}
\includegraphics{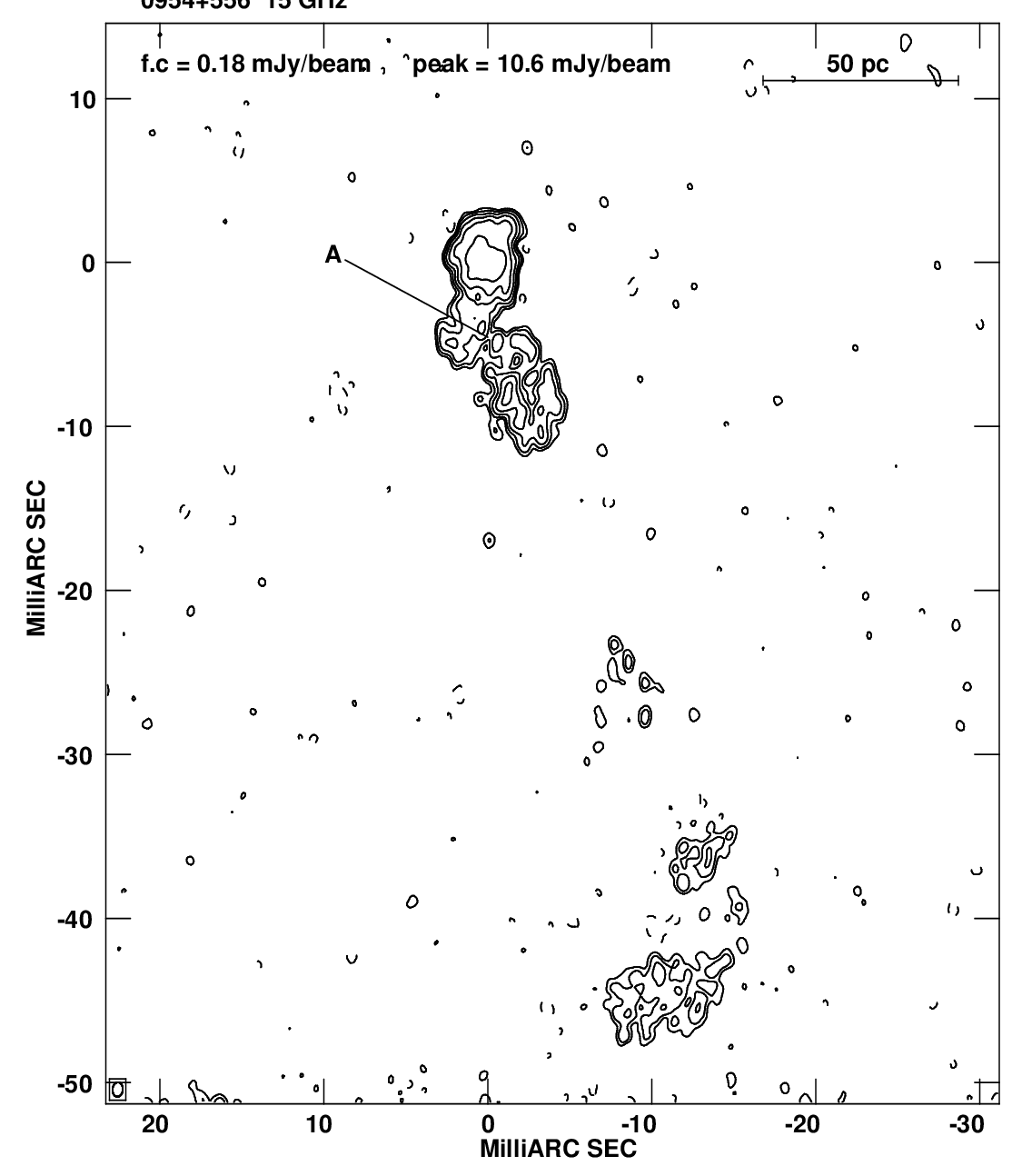} 
\vspace {24.0cm}  
\caption{continued.}
\end{center}
\end{figure}

\begin{figure}[htbp]
\begin{center}
\includegraphics{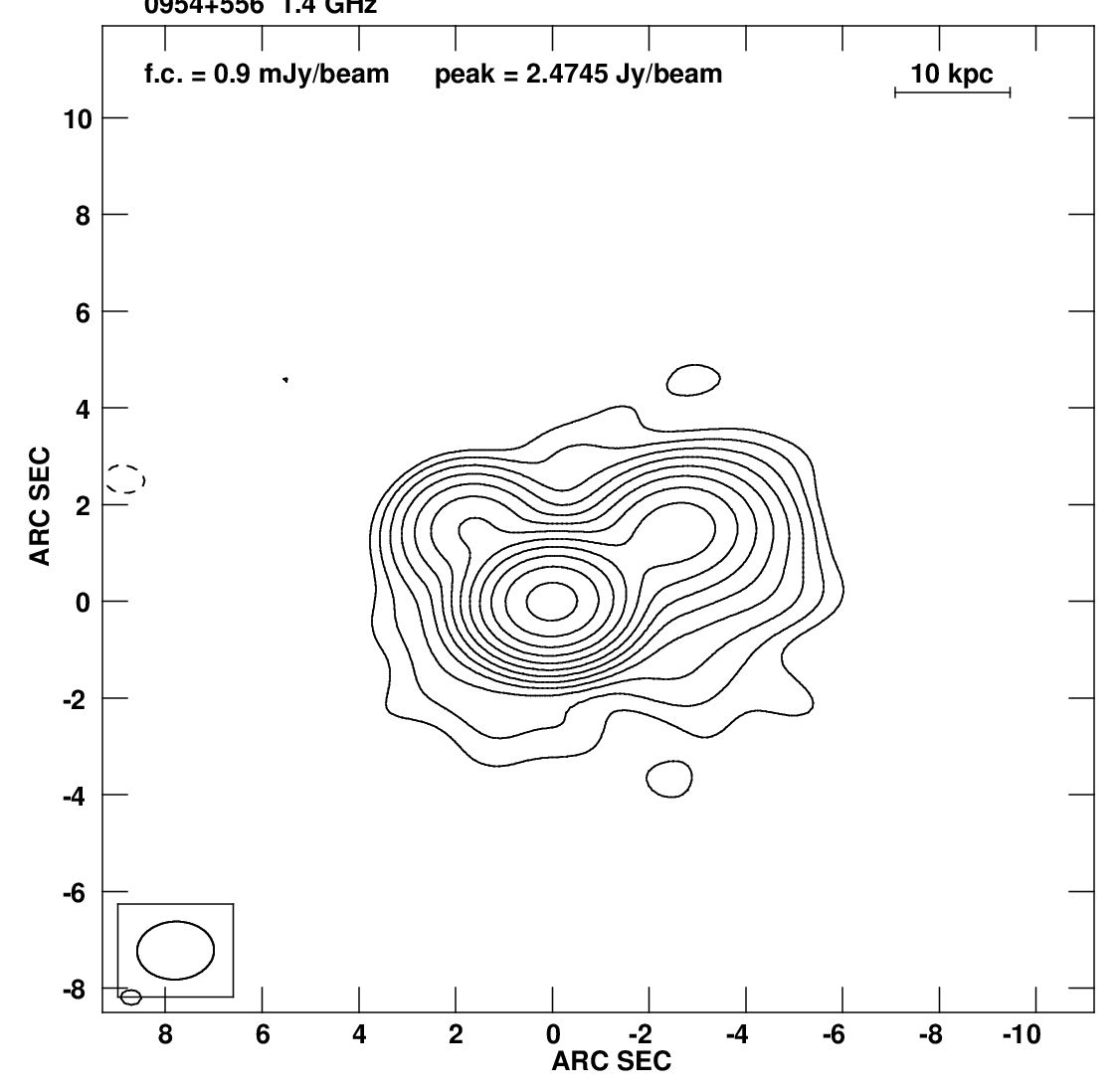}
\vspace {8.cm}
\caption{1.4~GHz VLA images of \object{0954+556}: the first contour (f.c.) is generally three times the rms noise level on the image; contour levels increase by a factor of 2; the restoring beam is shown in the bottom left corner of each image.} 
\label{0954}
\end{center}
\end{figure}

The South component ($B$) is characterized by a more diffuse emission, 
with $\alpha_{1.7}^{4.8}$=0.49. 
The overall structure of this object resembles that of \object{4C31.04} 
(Giroletti et al. \cite{Giroletti03}), one of the weakest known CSOs.
We would classify \object{0954+556} as a CSO, 
even if in our images there is no indication 
of a feature which could be the core candidate at a level $\gtsim 2$~mJy/beam.
Its total extent is $\sim 80$~mas, corresponding
to $\sim340$~pc.
However, VLA images at several frequencies (Hong et al. in preparation) show 
on arcsecond scale an
unresolved component containing the structure that we see on the tenth of
arcsec scale, and two further components extending out
for $2\farcs5$ to North-East and $4\farcs0$ to North-West respectively.
The overall size is $\approx 4\farcs5$, corresponding to a linear size of $\approx 19$~kpc,
qualifying the arcsecond radio structure as an MSO.
The missing flux density in our images is accounted for by the extended emission on
arcsecond scales.
In Fig.~\ref{0954} we show a VLA image at 1.4~GHz, with a resolution of $\sim 1\farcs4$,
that we have produced from VLA archive data, showing the overall 
structure. Note how much this source recalls 3C286 (Cotton et al.
\cite{Cotton97}).

The relation of the mas to the arcsec structure is at the moment far
from being understood.
It could be a case of a \emph{smothered} source (Baum et al. \cite{Baum90}) or of 
recurrent activity, although neither of these possibilities appears convincing.
Further investigation of this source is necessary. 

\begin{table*}[htbp]
\centering
 \caption{General parameters of PW CSOs/MSOs:  
Column 1: Source name;
Column 2: Optical identification (Id), G = galaxy, Q = quasar, EF= empty field;
Column 3: Redshift (z);
Column 4: Flux density (Jy) at 2.7~GHz, from PW;
Column 5: Spectral index $\alpha^{2.7}_{5.0}$, from PW;
Column 6: Maximum angular size (LAS);
Column 7: Frequency of spectral flux density peak;
Column 8: Linear Size (LLS);
Column 9: Log Radio power at 2.7~GHz;
Column 10: Concise structure description: sD, aD, dD, xD  = {\bf s}ymmetric, 
{\bf a}symmetric, {\bf d}istorted, comple{\bf x} {\bf D}ouble; c, J = 
{\bf c}ore, {\bf J}et(s) present; T = {\bf T}riple, i.e. three  
components without an obvious source center; Cx = {\bf C}omple{\bf x}. 
The discrimination between sD and aD is qualitative and is mostly based on the 
source arm length ratio (if a core is detected) or on the lobe flux density ratio
Column 11: Representative references for radio structure.}
\label{list}
\begin{tabular}{cllllllllllc}
 \hline
 \hline
Source            & Id  &   z  &$S_{2.7}$&$\alpha$&  LAS &$\nu_{max}$&  LLS  &log$P_{2.7}$& alternative   & Radio       & References   \\
                  &     &      &    Jy   &        &arcsec&   GHz     &  kpc  &   W/Hz     &  name         & Structure    &             \\
  (1)             & (2) &  (3) &   (4)   &    (5) & (6)  &   (7)     &  (8)  &   (9)      &     (10)      &      (11)    &   (12)      \\
\hline
\object{0026+346} &  G  & 0.52 &   1.5   &  0.28  &  0.04&  1.31     &  0.15 &    26.59   &               &  sD,c?       & 14,17,18    \\

\object{0116+319} &  G  & 0.06 &   2.1   &  0.58  &  0.1 &  0.4      & 0.08  &    24.91   &     4C31.04   &  sD,c        &    27       \\

\object{0127+233} &  Q  & 1.46 &   1.7   &  0.70  &  2.6 & $\le 0.1$ & 11.1  &    27.65   &3C43, 4C23.06  &  dD,c,J      &     1,9     \\
  
\object{0138+136} &  G  & 0.62 &   1.5   &  0.77  &  1.0 &   0.12    &  3.85 &    26.84   &3C49, 4C13.10  &  aD,c        &     3,9   \\
 
\object{0221+276} &  G  & 0.31 &   1.7   &  0.99  &  2.5 & $\le 0.1$ &  7.01 &    26.30   &3C67, 4C27.08  & aD,c?        &     3,6   \\
 
\object{0223+341} &  Q  & 2.91 &   1.8   &  0.53  &  1.1 &   0.15    &  4.04 &    28.18   & 4C 34.07      & aD           &   2,23,25   \\
 
\object{0316+162} &  G? & 1.0  &   4.9   &  0.89  &  0.3 &   0.8     &  1.28 &    27.83   & 4C 16.09, CTA21 &   aD,c     &   2,25      \\
 
\object{0404+768} &  G  & 0.60 &   3.5   &  0.58  &  0.15&   0.6     &  0.57 &    27.14   & 4C 76.03      & aD,c,J       &   2,4,5,25  \\
 
\object{0428+205} &  G  & 0.22 &   3.2   &  0.50  &  0.2 &   0.55    &  0.45 &    26.22   &               &  aD,c,J      &   2,25      \\
 
 
\object{0538+498} &  Q  & 0.55 &  13.0   &  0.75  &  0.7 &   0.35    &  2.58 &    27.67   &3C147, 4C49.14 &  aD,c,J      &   9,11,12  \\
 
\object{0710+439} &  G  & 0.52 &   1.9   &  0.23  & 0.024&   1.9     &  0.09 &    26.69   &               &  aD,c,J      &   4,5,26    \\
 
\object{0945+664} &  G  & 1.0  &   1.6   &  0.46  &  1.5 &  0.1      &  6.39 &    27.21   & 4C 66.09      &   sD?        &0,24   \\

\object{0954+556} &  Q  & 0.90 &   2.6   &  0.22  &  4.7 &$\le$ 0.1  &  19.7 &    27.26   & 4C 55.17      &  xD          &0   \\
 
 
\object{1031+567} &  G  & 0.45 &   1.58  &  0.30  & 0.031&   1.3     &  0.11 &    26.50   &               &  sD?         &   5,23      \\
 
\object{1153+317} &  Q  & 0.42 &   1.7   &  0.92  & 0.9  &   0.1     &  2.96 &    26.57   & 4C 31.38      &  sD          &   20        \\
 
\object{1203+643} &  G  & 0.37 &   2.0   &  0.97  &  1.3 &   0.08    &  4.02 &    26.52   &3C268.3, 4C64.14& aD,c        &     3,9    \\
  
\object{1225+368} &  Q  & 1.97 &   1.6   &  1.17  & 0.055&   1.0     &  0.23 &    28.11   &              &  aD,c,J       &     2,25    \\
 
\object{1250+568} &  Q  & 0.32 &   1.5   &  0.61  &  1.6 & $<$ 0.1   &  4.57 &    26.23   &3C277.1, 4C56.20&  aD,c,J     &     6,9    \\
 
\object{1323+321} &  G  & 0.37 &   3.3   &  0.58  & 0.06 &   0.35    &  0.19 &    26.69   & 4C 32.44     &   sD          &     2,14,15 \\
   
\object{1328+307} &  Q  & 0.85 &  10.3   &  0.51  &  3.8 &   0.1     &  15.8 &    27.90   &3C286, 4C30.26 &  xD,c        &   7,21      \\


\object{1358+624} &  G  & 0.43 &   2.7   &  0.66  & 0.05 &   0.5     &  0.17 &    26.75   & 4C 62.22      &  aD,c,J      &     2,4,25  \\

\object{1404+286} &  G  & 0.08 &   1.9   & -0.78  & 0.01 &   6.0     &  0.01 &    25.07   & OQ 208        &   aD,c       &   13,14,15  \\
 
\object{1413+349} & EF  & 1.0  &   1.7   &  0.60  & 0.06 &   0.8     &  0.26 &    27.28   &               &  aD,c,J      &   2,25      \\
 
 
\object{1458+718} &  Q  & 0.91 &   5.3   &  0.56  &  2.2 &   0.07    &  9.26 &    27.68   &3C309.1, 4C71.15& aD.c.J      &     7,10,17 \\
 
\object{1607+268} &  G  & 0.47 &   2.9   &  0.89  & 0.055&   1.1     &  0.19 &    26.90   &CTD 93          &   sD        &     17,22,19\\
 
\object{1637+626} &  G  & 0.75 &   2.2   &  1.00  & 0.38 &   0.3     &  1.54 &    27.26   &3C343.1, 4C62.27&   sD        &   3,9      \\
 
\object{1819+396} &  G  & 0.80 &   1.8   &  0.98  & 1.0  &    0.2   &  4.11 &     27.20   & 4C 39.56     &  dD,c?        &   2,20,24  \\
 
\object{1829+290} &  G  & 0.84 &   1.9   &  0.80  & 3.0  & $<$0.1    & 12.5 &     27.23   & 4C 29.56     &   dD          &   2,20     \\
 
\object{2021+614} &  G  & 0.23 &   2.2   & -0.10  & 0.011&  2.82     &  0.03 &    26.04   &              &  sD,c         &  14,15,16  \\

\object{2252+129} &  Q  & 0.54 &   1.5   &  0.75  & 4.0  & $<$ 0.05  & 14.7 &     26.71   &3C455, 4C12.79 &  sD,c?,J?    &   7,8,20   \\
 
\object{2342+821} &  G  & 0.74 &   2.3   &  0.92  & 0.18 &    0.5    &  0.73 &    27.22   &              &      T        &   2,25     \\
 
\object{2352+495} &  G  & 0.24 &   2.18  &  0.34  & 0.05 &    0.7    &  0.12 &    26.11   &              &   dD,c,J?     &   4,5,26   \\
 
   \hline\hline
\object{0134+321} &  Q  & 0.37 &   8.97  &  0.83  & 1.5  &    0.08   &  4.64 &    27.16   &    3C48      &    Cx          & 30,31 \\

\object{1328+254} &  Q  & 1.06 &   4.6   &  0.63  & 0.1  &    0.05   &  0.43 &    27.77   &    3C287     &    Cx          & 29    \\

\object{1600+335} & G?  & 1.1  &   2.2   &  0.63  &$<0.1$&    0.95   &$<0.43$&    27.48   &              &    Cx          & 20,25 \\

\object{1634+628} &  Q  & 0.99 &   2.7   &  0.95  & 0.25 &    0.25   &  1.06 &    27.57   &    3C343     &    Cx          &  3,28    \\
\hline

\end{tabular}
\vspace {0.15cm}
\begin{list}{}
\item 
(0) this paper;
(1) Fanti et al. (\cite{Fanti02}); 
(2) Dallacasa et al. (\cite{Dallacasa95});
(3) Fanti et al. (\cite{Fanti85});
(4) Readhead et al. (\cite{Readhead96a});
(5) Taylor et al. (\cite{Taylor96});
(6) Sanghera et al. (\cite{Sanghera95});
(7) Akujor \& Garrington (\cite{Akujor95});
(8) Bogers et al. (\cite{Bogers94});
(9) L\udier dke et al. (\cite{Ludke98});
(10) Wilkinson et al. (\cite{Wilkinson86});
(11) Junor et al. (\cite{Junor99});
(12) Nan et al. (\cite{Nan00});
(13) Stanghellini et al. (\cite{Stanghellini97});
(14) Kellermann et al. (\cite{Kellermann98});
(15) Fey et al. (\cite{Fey96});
(16) Tschager et al. (\cite{Tschager99});
(17) Fey et al. (\cite{Fey97});
(18) Taylor et al. (\cite{Taylor94});
(19) Stanghellini et al. (\cite{Stangh02});
(20) Spencer et al. (\cite{Spencer89})
(21) Cotton et al. (\cite{Cotton97});
(22) Shaffer et al. (\cite{Shaff99});
(23) Fomalont et al. (\cite{Fom00});
(24) Xu et al. (\cite{Xu95});
(25) Dallacasa et al. (\cite{Dallacasa04});
(26) Taylor et al. (\cite{Taylor00});
(27) Giroletti et al. (\cite{Giroletti03});
(28) Ren Dong et al. (\cite{Rend88});
(29) Fanti et al. (\cite{Fanti89});
(30) Fanti et al. (\cite{Fanti90});
(31) Wilkinson et al. (\cite{Wilk91})
(32) Lister et al. (\cite{Lister03}).
\end{list}
\end{table*} 

\section{The list of CSOs \& MSOs from the PW catalogue}

In Table ~\ref{list} we give the list with relevant data compiled 
from the literature of all the ``compact'' sources from the PW
catalogue which, on the basis of literature data, {\it are not} genuine {\it
core--jet} radio sources. We are confident that all the CSOs/MSOs in the 
catalogue have been spotted and that the list is complete. 
Most of these sources can be classified as
CSOs/MSOs. A few of them have an ``anomalous'' complex  structure.  For 
completeness, we also list them in the bottom section of the Table~\ref{list}. 
In the PW sources \object{0831+55},
\object{3C216}, \object{1345+125}, \object{3C299}, \object{1502+10}, 
originally classified by the authors
as $U$ (unresolved, $<2$~arcsec), from subsequent more sensitive 
observations revealed additional low brightness emission. These objects 
are therefore extended radio sources.  

\section{Conclusions}
		     
We have presented the results of multi-frequency VLBA observations for 6 
sources selected from the PW catalogue on the basis of their ambiguous 
morphological and spectral classification. 

We confirm the core-jet structure of three of them (\object{0133+476}, \object{0153+744},
\object{0859+470}), all identified as quasars. Hints for superluminal structure changes are found 
from a comparison with literature data. Also components with no detectable 
motion over long time scales are present.

One source (\object{0202+149}) is, in many respects, an ``enigmatic'' object. Its radio
structure may be that of a core-jet or a very asymmetric CSO, although we feel that further investigations
are needed to firmly establish which is the radio core.

The last two objects are kpc scale MSOs. \object{0945+664} (whose suggested optical
identification is with a faint galaxy without red-shift) has a double 
structure of which we detect the bright hot spots. The core is not detected.
\object{0954+556} (a quasar) has a double structure on a
tens of mas scale ($\sim 340$~pc) plus arcsec (19~kpc) structure
at a position angle roughly perpendicular to that of the small scale. 
The core is not detected and the connection of the small scale to the
large scale structure is not understood.

We are confident that the last two sources complete the sample 
of CSOs/MSOs present in the PW catalogue
and the complete list is given in Table~\ref{list}.
This sample is very useful to study the radio source distribution in the 
``Radio Power -- Linear Size'' plane in the range of linear sizes from 
$\approx$ 10~pc to $\approx$ 20~kpc, which is an important tool to constrain
source-evolution models (Baldwin \cite{Baldwin82}).

\begin{acknowledgements}
The VLBA is operated by the U.S. National Radio Astronomy Observatory which is
a facility of the National Science Foundation operated under a cooperative
agreement by Associated Universities, Inc. 
This research has made use of the NASA/IPAC Extragalactic Database (NED) which 
is operated by the Jet Propulsion Laboratory, California Institute of Technology, 
under contract with the National Aeronautics and Space Administration.  
This research has also made use of the United States Naval Observatory (USNO) 
Radio Reference Frame Image Database (RRFID) and
of data from the University of Michigan Radio Astronomy Observatory which has 
been supported by the University of Michigan and the National Science Foundation. This work was partly supported by the Italian Ministry for University and Research (MIUR) under grant COFIN 2002-02-8118.
\end{acknowledgements}

\end{document}